\documentclass[prd,twocolumn,superscriptaddress,floatfix,amsmath,amssymb,amsfonts,longbibliography,nofootinbib]{revtex4-1}
\usepackage{float} 

\usepackage[normalem]{ulem}
\usepackage[english]{babel}
\usepackage{graphicx}% Include figure files
\usepackage{dcolumn}% Align table columns on decimal point
\usepackage{bm}% bold math
\usepackage{blindtext}
\usepackage{verbatim}
\usepackage{mathrsfs}
\usepackage{amsmath}
\usepackage{cancel}
\usepackage{physics}
\usepackage{epstopdf}
\usepackage{mathtools}
\usepackage{tensor}
\usepackage{color}
\usepackage[usenames,dvipsnames]{pstricks}
\usepackage{epsfig}
\usepackage{pst-grad} % For gradients
\usepackage{pst-plot} % For axes
\usepackage{hyperref}
\usepackage{verbatim}
\usepackage{slashed}

\renewcommand{\H}{\mathcal{H}}
\newcommand{\tb}[1]{\textcolor{black}{#1}}

\renewcommand{\b}{\color{black}}%to see implemented changes change black -> blue in this command
\renewcommand{\c}{\color{black}}

\newcommand{\fix}[1]{\textcolor{black}{#1}}

\allowdisplaybreaks[1] %<<<<<<<<<<<<<<<<<<<<<<
\begin{document}

%\title{Schr\"odinger Wavefunctions in Curved Spacetimes}
\title{A Wavefunction Description for a Localized Quantum Particle in Curved Spacetimes}

\author{T. Rick Perche}
\email{trickperche@perimeterinstitute.ca}
\affiliation{Department of Applied Mathematics, University of Waterloo, Waterloo, Ontario, N2L 3G1, Canada}
\affiliation{Perimeter Institute for Theoretical Physics, Waterloo, Ontario, N2L 2Y5, Canada}

\author{Jonas Neuser}
\email{jonas.neuser@fau.de}

\affiliation{Institute for Quantum Gravity, University of Erlangen-Nürnberg, Staudtstraße 7 / B2, 91058 Erlangen, Germany}

\begin{abstract}
    \tb{%Schr\"odinger's equation for a complex wavefunction is intrinsically non-relativistic. This limits its applications to non-relativistic scenarios in flat spacetimes. On the other hand, Dirac's spinor formalism is fully relativistic and well understood within the framework of general relativity. In this work 
    We reduce Dirac's spinor formalism for a spin $1/2$ particle to a complex wavefunction description in curved spacetimes. We consider a localized fermionic particle in curved spacetimes and perform an expansion in terms of the acceleration and curvature around the center of mass of the system, generalizing the results of \b \cite{ParkerRevLett,parker}\c. Under a  non-relativistic approximation, one obtains a quantum description in a Hilbert space of complex wavefunctions defined in the rest space of the system. The wavefunction of the particle then evolves according to a modified Schr\"odinger equation associated with a symmetric Hamiltonian. When compared to the standard Schr\"odinger equation for a wavefunction, we obtain corrections in terms of the acceleration of the system's center of mass and curvature of spacetime along its trajectory. In summary, we provide a formalism for the use of a complex wavefunction to describe a localized quantum particle in curved spacetimes.}
\end{abstract}

\maketitle

\section{Introduction} 
    \b
    Although physics is often concerned with descriptions of processes in terms of more fundamental concepts, in order to describe more complex systems, it is usual to employ an effective and less fundamental formulation. For instance, quantum field theory (QFT) is among the most successful physical theories ever formulated~\cite{weinberg,Peskin}. In fact, the theory is even well defined in relativistic contexts and is compatible with Einstein's general relativity in the semiclassical regime~\cite{birrell_davies,Wald2}. However, there are still issues within QFT that require a deeper understanding. For instance, although non-perturbative processes are well understood in specific examples~\cite{nonPerturb2D,nonPerturb}, there is no general approach for non-perturbative problems in QFT. In fact, when describing localized bound states in quantum electrodynamics the mechanism for binding the electron is still regarded as a classical electromagnetic field~\cite{boundQED,boundQEDEx}. Another issue is the definition of projective measurements within the context of quantum field theory~\cite{Sorkin,measurmentEdu}, which have been shown to violate causality. In fact, one way to provide a consistent measurement theory for a quantum field is by coupling a localized non-relativistic probe to it~\cite{measQFT}.

    In summary, although we have fundamental quantum field theories that describe some physical setups very precisely, we still require other less fundamental theories in order to describe numerous physical situations, such as bound electron systems and non-relativistic probes of quantum fields~\cite{us,delocAtom}. {One of the physical scenarios where these theories are relevant is when considering the Unruh effect}~\cite{Unruh1976,fullingUnruhEffect,Davies1974}. In fact, it has been argued that uniformly accelerated atoms can in principle be used as probes of Unruh-like effects~\cite{scullyPRL,unruhSlow,vanzella,accAtomsCavity,Hu}. Simpler theories that describe bound states can also be relevant when probing Hawking radiation \cite{radiation,HawkingRadiation}, where the effect of gravity in the system might play a crucial role. Moreover, there are recent experiments that use atoms to probe weak gravitational fields~\cite{prlInterferometer,exp1,exp2} and there has also been recent interest in investigating quantum gravitational effects \cite{gravDetec,remi} by means of atomic probes. 
    
    Among the first studies of the interaction of atoms with gravity are the works of Leonard Parker~\cite{parker,ParkerRevLett,ParkerHydrogen}, where the energy level corrections of a geodesic atom under the influence of spacetime curvature were computed. Parker's works provided a framework for exploring the coupling of unaccelerated atoms with gravitational fields by treating the electron in an atom as a Dirac spinor. However, given that there has been growing interest in the description of accelerated probes \cite{accHydro,ZhaoHydrogen,ZhaoHydrogen2}, it is only natural to seek an extension of the works \cite{parker,ParkerRevLett,ParkerHydrogen} that also takes into account acceleration. We extend the works of Parker by also considering arbitrary states of motion for the system, thus obtaining a framework able to describe arbitrarily accelerated atoms \cite{accHydro,ZhaoHydrogen,ZhaoHydrogen2} and also takes into account the effect of gravity.
    
    We are then concerned with the description of a one-particle fermionic state localized around an arbitrary trajectory in curved spacetimes. Problems that involve a single fermion can be described by a spinor field that satisfies Dirac's equation in a curved background, provided that the energy of the system is small compared to the particle's mass%\footnote{\b If the system's energy surpasses twice the particle's mass, then particle production starts playing a role and a single particle description is not appropriate.\c}
    ~\cite{guideForCalculations}. However, there are very limited cases of analytical solutions to Dirac's equation even in flat spacetimes and yet less in curved backgrounds~\cite{guideForCalculations,solutions}. For this reason, rather than a spinor formalism, we seek an approximate description for one-particle systems in terms of complex wavefunctions that satisfy a modified Schr\"odinger equation. This description also provides a formalism which is compatible with theories that couple localized systems to quantum fields, where the probe is usually considered to be a non-relativistic quantum system~\cite{eduardo,us,us2,remi}.

    \c

    In this work we present the reduction of Dirac's spinor description to Schr\"odinger's wavefunction formalism for \tb{a} localized \tb{one-particle} system in \tb{its} proper frame. Our results are given in terms of an expansion to first order in the acceleration of the system's worldline and curvature of spacetime along its trajectory. This description \tb{allows one to identify the Hilbert space associated with the description of the particle and} provides a purpose to Schr\"odinger's complex wavefunction formalism in general relativistic scenarios. \b The expansion is performed in the reference frame of the center of mass of the system, using the Fermi normal coordinates (FNC) around its worldline. The major result of this manuscript is the generalized Schr\"odinger Hamiltonian \eqref{YES}, which {provides a} complex wavefunction {description for a particle} in its center of mass frame{,} with corrections due to curvature and acceleration. \c

    This manuscript is organized as follows: in Section \ref{secSchAsDir} we review the calculations that reduce Dirac's description \tb{for a spin 1/2 particle} to the Schr\"odinger \tb{formalism} in \tb{the case of} inertial reference frames in Minkowski spacetime. Section \ref{secSpniCur} is devoted to the review of the most relevant properties of spinors in curved spacetimes \tb{and setting the conventions that will be used throughout the manuscript}. In Section \ref{secSynge} we present the mathematical apparatus associated with Synge's world function and discuss the FNC around a \tb{timelike} curve. We also review the expansions of the metric and Christoffel symbols in these coordinates in terms of curvature and acceleration. In Section \ref{secDirWorldline} we expand Dirac's equation in terms of acceleration and curvature along \tb{a particle's}  worldline, obtaining an effective Hamiltonian for the system. In Section \ref{secSchInCur} we present our main results regarding the reduction of Dirac's theory \tb{for a spin 1/2 particle} to Schr\"odinger's wavefunction description in curved spacetimes and identify the suitable \tb{Hilbert space and} observables of the theory. In Section \ref{secAtomGravity} we apply our formalism to the example of a static atom under the influence of gravity, and a fermion in AdS spacetime. The conclusions of our work can be found in Section \ref{secConclusion}. 
    \b
    This manuscript contains four appendices, regarding four explicit computations. We compute the spin connection associated with the Fermi normal coordinates around the system's world line in Appendix \ref{appFrame}, the Dirac Hamiltonian in these coordinates in Appendix \ref{appDirH} and the reduction of Dirac's Hamiltonian to act on wavefunctions in Appendix \ref{appSchCur}. Finally, in Appendix \ref{appHermitian} we show that the obtained Hamiltonian is Hermitian with respect to the adequate inner product.

    In this manuscript we work with natural units such that $c = \hbar = 1$.% and choose the East Coast (mostly plus) convention for the spacetime metric.
    \c

\section{Schr\"odinger Equation as a Limit of Dirac Equation}\label{secSchAsDir}
    
    In order to be as self-contained as possible, we briefly review the concept of spinors in Minkowski spacetime and the reduction of Dirac's equation to Schr\"odinger's under a nonrelativistic approximation \b\cite{Wald1,sakurai,Shankar}. This reduction provides a simplified treatment for quantum systems in terms of complex wavefunctions. The resulting theory successfully describes multiple non-relativistic scenarios, such as the Schr\"odinger atom and electrons in most potentials \cite{cohen,sakurai}. Later, in Section \ref{secSchInCur}, we will perform a similar reduction in the context of systems localized around arbitrary trajectories in curved spacetimes. \c

    In special relativity it is crucial to keep track of the way objects transform under actions of the Lorentz group. In this work, we will regard a spinor as an object that transforms according to the $(\frac{1}{2},0)\oplus(0,\frac{1}{2})$ representation of the Lorentz algebra. Thus, a spinor is an element of a $4$ dimensional complex vector space. A spinor field   
    %One example are vector fields, which is four-dimensional. But there is yet another interesting four dimensional representation, the Dirac spinor.
    %Once upon a time 
    %A spinor field on Minkowski spacetime is an infinite dimensional representation for the Lorentz group. A spinor is an object that lives in a space where Lorentz transformations  act according to the $SL(2,\mathbb{C})$ representation.
    %A spinor $\psi(x)$ can be seen as an object that transforms under the spin representation of the Lorentz group. Using the $(\frac{1}{2},\frac{1}{2})$ representation of the Lorentz group in Minkowski spacetime,
    can then be written as
    \begin{equation}\label{1}
        \psi = \begin{pmatrix}\psi_1\\\psi_2\\\psi_3\\\psi_4\end{pmatrix} = \begin{pmatrix} \psi_A \\ \psi_B\end{pmatrix},
    \end{equation}
    where $\psi_1,\psi_2,\psi_3,\psi_4$ are complex functions \tb{in spacetime} and $\psi_A,\psi_B$ are 2 dimensional complex vectors. In this section we assume to have inertial coordinates $x = (t,\bm x)$.
    
    Equivalently, the spinor space can be defined as a fundamental representation of the Clifford algebra. The Clifford algebra in Minkowski spacetime requires $4$ generators, that are usually denoted $\gamma^\mu$, $\mu = (0,1,2,3)$. These are called the gamma matrices and satisfy the anticommutation relations $\{\gamma^\mu, \gamma^\nu\} = - 2 \eta^{\mu \nu}\b\openone_4\c$, where we use the convention $\eta_{\mu\nu} = \textrm{diag}(-1,1,1,1)$\footnote{\textcolor{black}{Notice that in order to use the standard convention for the $\gamma^\mu$ matrices and the mostly plus metric convention, we add a minus sign to the Clifford algebra anticommutation relations.}} for the Minkowski metric. In our description we use the Dirac basis, so that the gamma matrices $\gamma^\mu$ take the following form
    \begin{align}
        \gamma^0 &= \begin{pmatrix}
        \b\openone_2\c & \\ & -\b\openone_2\c
        \end{pmatrix},&&&
        \gamma^i &= \begin{pmatrix}
              & \sigma^i  \\
             -\sigma^i & 
        \end{pmatrix}.
    \end{align}

    The Dirac Lagrangian for a spinor coupled to an external electromagnetic field $A_\mu$ can then be written as
    \begin{equation}
        \mathcal{L} = \frac{1}{2}\bar{\psi}(i\slashed \partial +q\slashed A - m)\psi+\textrm{H.c.},
    \end{equation}
    where $q$ is the electromagnetic charge of the spinor field and we use Feynman's slash notation $\slashed b = \gamma^\mu b_\mu$ \b and $\textrm{H.c.}$ denotes the Hermitian conjugate\c. The extremization of the associated action yields Dirac's equation as an equation of motion for the spinor $\psi$:
    \begin{equation}\label{diracFlat}
        (i\slashed \partial + q\slashed A- m)\psi = 0.
    \end{equation}
    
    Having Dirac's equation associated with a given inertial reference frame allows one to split the time and space components of the equation and associate a  Hamiltonian to the system. We obtain
    \begin{align}\label{5}
        %i\gamma^0 \partial_t \psi = -i \gamma^i \partial_i \psi + m \psi\\
        i \partial_t \psi = \gamma^0\left(-i \gamma^i \partial_i-q\slashed A + m\right)\psi,
    \end{align}
    where we have used $(\gamma^0)^2 = \openone$ to isolate the generator of time translations with respect to the frame. One then identifies the Hamiltonian operator for the spinor system as
    \begin{equation}\label{HSpinorFlat}
        H = \gamma^0\left(-i \gamma^i \partial_i-q \slashed A + m\right).
    \end{equation}

    \tb{Dirac's formalism can be used to describe a spin 1/2 quantum particle. Indeed, the Hilbert space associated with the underlying quantum theory can be seen as the space of spinors defined at a given instant of time equipped with the following inner product,
    \begin{equation}\label{innSpinFlat}
        (\phi,\psi) = \int \dd^3 x \:\phi^\dagger(x)\psi(x).
    \end{equation}
    Elements of this Hilbert space then evolve according to the Hamiltonian of Eq. \eqref{HSpinorFlat}. Expected values of observables of the theory can also be computed using standard techniques of quantum mechanics.} %For instance, the expected value of the momentum $p_i$ of the Dirac particle is given by $(\psi,-i\partial_i\psi)$.

    To reduce Dirac's formulation to Schr\"odinger's, we use the decomposition of $\psi$ in terms of the two-component spinors $\psi_A$ and $\psi_B$ from Eq. \eqref{1}. Equation \eqref{5} then results the following equations
    \begin{align}
        i\partial_t \psi_A = H_{AA}\psi_A + H_{AB}\psi_B,\label{psiAeqn}\\
        i \partial_t \psi_B = H_{BA}\psi_A+H_{BB}\psi_B,\label{psiBeqn}
    \end{align}
    where %$H_{IJ} = P_I H P_J$ for $I,J = A,B$ and $P_A,P_B$ are the projectors on the spaces $A$ and $B$. Explicitly, we have 
    \begin{align}
        H_{AA} &= - qA_0+m,
        &&& H_{AB} &= -(i \sigma^i\partial_i + q \sigma^i A_i),\\
         H_{BB} &= -qA_0 -m, &&&H_{BA} &=  -(i \sigma^i\partial_i + q \sigma^i A_i).\label{8}
    \end{align}
    Notice that the fact that Dirac's Hamiltonian is Hermitian implies $H_{AB}=H_{BA}$.
    
    To obtain Schr\"odinger's equation, one usually proceeds by solving the equation of motion for the $B$ component of the spinor in terms of the $A$ component. Equation \b \eqref{psiBeqn} \c for $\psi_B$ reads
    \begin{equation}
        -(i \partial_t + qA_0 +m)\psi_B = i\sigma^i(  \partial_i- iq A_i) \psi_A.
    \end{equation}
    In order to perform a nonrelativistic expansion, we identify the nonrelativistic energy operator as the total energy of the system \b minus \c its rest mass,
    \begin{equation}
        i\partial_T = i \partial_t -m.
    \end{equation} 
    With this we can rewrite $i \partial_t +m = i \partial_T +2m$. In the nonrelativistic regime, the expected value of the nonrelativistic energy $i\partial_T$ and the electromagnetic potential is much smaller than the rest mass of the particle. This allows us to perform the following expansion
    \begin{align}
        \psi_B &= - (2m+qA_0+i\partial_T)^{-1}i \sigma^i (  \partial_i- iq A_i) \psi_A\nonumber\\
        &=-\frac{1}{2m} \sum_{n=0}^\infty \b (-1)^n\!\c\left(\frac{i\partial_T+qA_0}{2m}\right)^n \! i \sigma^i(  \partial_i- iq A_i) \psi_A\nonumber\\
        &\approx \frac{1}{2m i} \sigma^i(  \partial_i- iq A_i) \psi_A\equiv D_B\psi_A,\label{approxFlat}
    \end{align}
    where have only carried the $0$th order term and $D_B$ is defined by Eq. \eqref{approxFlat}.
    
    Plugging the result for $\psi_B$ into Eq. \eqref{psiAeqn} we obtain 
    \begin{align}
         i\partial_t \psi_A =& (H_{AA} + H_{AB}D_B)\psi_A\\
         =&- \frac{1}{2m }\sigma^i\sigma^j  (  \partial_i- iq A_i)(  \partial_j- iq A_j) \psi_A\\& \phantom{00000000000}+ m \psi_A -qA_0\psi_A\nonumber
    \end{align}
    Then we use the following relation for the product of Pauli matrices:
    \begin{equation}\label{traceProp}
        \sigma^i\sigma^j = \delta^{ij} + i \epsilon\indices{^{ij}_k} \sigma^k,
    \end{equation}
    \b where $\epsilon_{ijk}$ is the Levi-Civita symbol and we raise Latin indices with the Kronecker delta. This yields\c
    \begin{align}
         i\partial_T \psi_A =& - \frac{1}{2m }(  \partial^i- iq A^i)(  \partial_i- iq A_i)  \psi_A\nonumber\\&-qA_0 \psi_A-\frac{q}{m} \epsilon\indices{^{ij}_k} \sigma^k \partial_iA_j\psi_A,\label{diracFlatA}
    \end{align}
    where we have again used the definition of the nonrelativistic energy operator $i\partial_T$. Equation \eqref{diracFlatA} represents the equation of motion for a two-component complex-valued wavefunction. This is the formalism commonly employed to study the fine structure of the hydrogen atom due to the spin-orbit term (last term in Eq. \eqref{diracFlatA}). The last step to obtain an equation for a complex scalar function $\psi(x)$ is to take the trace of the effective Hamiltonian $H_{AA}+H_{AB}D_B$, over the spin degrees of freedom. This yields an equation for a complex scalar function,
    \begin{equation}\label{Schrodinger}
        i\partial_T \psi= -q A_0 \psi- \frac{1}{2m }(  \partial^i- iq A^i)(\partial_i- iq A_i)\psi ,
    \end{equation}
    which is exactly the Schr\"odinger equation for a free particle \tb{of mass $m$} coupled to an external electromagnetic field. 
    
    \tb{Under this nonrelativistic approximation and the trace over spin degrees of freedom, the inner product from Eq. \eqref{innSpinFlat} reduces to the standard $L^2(\mathbb{R}^3)$ inner product in each space slice. That is, given two complex wavefunctions $\phi$ and $\psi$, we obtain
    \begin{equation}
        (\phi,\psi) = \int \dd^3x \:\phi^*(x)\psi(x).
    \end{equation}
    One can then identify the Hilbert space associated with the underlying quantum theory that describes a nonrelativistic particle as the space of square integrable wavefunctions for each value of time.} %\textcolor{red}{If we want to say something about the inner product here we will have to say something about the inner product for spinors previously. I have a commented paragraph on that, if you want to check.}
    
    Notice that to reach \tb{the} expression \tb{from Eq. \eqref{Schrodinger}} we needed to perform the expansion of Eq. \eqref{approxFlat}, which is intrinsically associated with the spacetime decomposition in the reference frame considered. Therefore, this expansion allows us to treat a nonrelativistic spinor as a complex wavefunction in the coordinates $(t,\bm x)$. Our goal now will be to generalize this approach to accelerated trajectories in curved spacetimes. In order to do so, we will first describe spinors in curved spacetimes. The next step will be to fix an explicit spacetime decomposition associated with the system. Finally, in Section \ref{secSchInCur}, we will perform a similar expansion to Eq. \eqref{approxFlat}.

\section{Spinors in Curved Spacetimes}\label{secSpniCur}
    
    \b
    In the previous section we have seen how to reduce Dirac's flat spacetime spinor description to a complex wavefunction formalism associated to an inertial frame. Nevertheless, Dirac's equation can also be used to describe a fermionic particle in the context of general relativity~\cite{stewartSpinors,Wald1,penroseSpinors}. The goal of this manuscript is to generalize the reduction done in Section \ref{secSchAsDir} to the context of a fermionic system localized around an arbitrary trajectory in curved spacetimes. For this purpose, \c  in this section we review the formalism of spinors in four dimensional Lorentzian manifolds \b\cite{guideForCalculations,stewartSpinors,Wald1,penroseSpinors} \c and fix the conventions that will be used throughout the manuscript. 
    
    \b
    The description of spinors in a curved spacetime $\mathcal{M}$ relies on defining a spin structure, and thus the spinor bundle, $S\mathcal{M}$. However, due to topological intricacies, not every spacetime admits a spin structure~\cite{penroseSpinors,Wald1}. In order to define spinors in a consistent manner, we will assume our spacetime to be time oriented and spacetime oriented. On top of these conditions, we will also assume that there exists a globally defined orthonormal frame over $\mathcal{M}$ so that Geroch's theorem ensures that a spinor bundle can be built~\cite{penroseSpinors,Wald1}.
    
    The spinor bundle $S\mathcal{M}$ is an associated bundle to an $SL(2,\mathbb{C})$ principal bundle and can be built from the tangent bundle $T\mathcal{M}$. In order to do that, given that $\mathcal{M}$ is spacetime orientable, we can write $T\mathcal{M}$ as an associated bundle to an $SO(1,3)$ principal bundle. Using the fact that $SL(2,\mathbb{C})$ is the universal (double) cover of $SO(1,3)$ allows us to build the spinor bundle from $T\mathcal{M}$ and import a natural connection from it. This will be essential in order to describe Dirac's equation in curved spacetimes.
    
    We will now explicitly build the spinor bundle and the spin connection from a global frame in $T\mathcal{M}$. \c 
    Let $(e_I)$, with ${I = (0,\mathrm{i})}$, be any orthonormal frame on the tangent bundle. We use the convention that capitalized Latin indices are associated with this frame and the space components are denoted by non capitalized, non italicized indices, e.g. $J = (0,\textrm{j})$, not to be confused with $(0,j)$. The orthonormality of the frame ensures that
    %The next step is to induce a connection in the spinor bundle from the metric compatible connection in the tangent bundle. There are different equivalent ways to do this, but the method we will choose here will be associated with an orthonormal frame in $T\mathcal{M}$. Let $(e_I)$ be any orthonormal frame in the tangent bundle, so that
    \begin{equation}
        e_I^\mu e_J^\nu g_{\mu\nu} = \eta_{IJ}.
    \end{equation}
    The connection 1-form associated with this frame, $\omega\indices{_\mu^J_I}$ is defined by
    \begin{equation}
        \nabla_\mu e_I = \omega\indices{_\mu^J_I}e_J.
    \end{equation}
    %and the orthonormality of the frame implies antisymmetry of the connection,
    %\begin{align}
        %\omega_{\mu IJ} = - \omega_{\mu J I},&& \textrm{where} &&\omega_{\mu I J} =  \omega\indices{_\mu^K_J}\eta_{IK}.
    %\end{align}
    %With this, the covariant derivative of any vector field $X^\mu$ can be written as
    %\begin{equation}
        %\nabla_\mu X^I = \partial_\mu X^I + \omega\indices{_\mu^I_J}X^J.
    %\end{equation}
    %This can now be rewritten in terms of the generators of the Lorentz group $J^{IJ}$as
    %\begin{equation}
        %\nabla_\mu X^I = \partial_\mu X^I + \frac{1}{2}\omega_{\mu KL} (J^{KL})\indices{^I_J}X^J.
    %\end{equation}
    %The expression above translates to an explicit basis the notion that the connection is defined in the level of the Lie algebra. This means that a natural extension of the covariant derivative $\nabla$ to the spinor bundle is given by
    %$\gamma\indices{^{(I|a|}_c}\gamma\indices{^{J)c}_b} = -\eta^{IJ}\delta^a_b$
    The generators of the Clifford algebra can be encoded into a tensor $\gamma$ of rank $(1,0)$ in $T\mathcal{M}$ and $(1,1)$ in $S\mathcal{M}$ such that $\{\gamma^I,\gamma^J\} = -2\eta^{IJ}\b\openone_4\c$.
    We introduce a frame $(E_a)$, $a=1,2,3,4$ in $S\mathcal{M}$ such that the components of $\gamma$ are given by
    \begin{align}
        \gamma^0 &= \begin{pmatrix}
        \b\openone_2\c & \\ & -\b\openone_2\c
        \end{pmatrix},&&&
        \gamma^\textrm{i} &= \begin{pmatrix}
              & \sigma^\textrm{i}  \\
             -\sigma^\textrm{i} & 
        \end{pmatrix}.
    \end{align}
    The frame $(E_a)$ is analogous to the Dirac basis for spinors in flat spacetimes. We fix the convention that the first letters of the Latin alphabet \b are \c used to denote indices associated with this frame.
    
    The covariant derivative can be extended to act on spinors according to
    \begin{equation}\label{covDSpinor}
        \nabla_\mu\psi^a = \partial_\mu \psi^a + \Gamma\indices{_\mu^a_b} \psi^b.
    \end{equation}
    \b Notice that the equation above contains two types of indices: spinor indices $a$ and $b$ that run from $1$ to $4$ (associated with the components of $\psi$) and a spacetime index $\mu$ that runs from $0$ to $3$ (associated with the covariant derivative). $\Gamma_\mu$ in Eq. \eqref{covDSpinor} is the spin connection~\cite{guideForCalculations}, explicitly given by\c
    \begin{equation}\label{ChristSpin}
        \Gamma\indices{_\mu^a_b} = -\frac{1}{2}\omega\indices{_{\mu I J}} (S^{IJ})\indices{^a_b},
    \end{equation}
    where we lower the indices of the connection 1-form with the Minkowski metric and $S^{IJ}$ are the generators of the $SL(2,\mathbb{C})$ action in the spinor bundle. These are given in terms of the $\gamma$ matrices by
    \begin{equation}
        S^{IJ} = \frac{1}{4}[\gamma^I,\gamma^J].
    \end{equation}
    
    With this we can then write down Dirac's Lagrangian $\mathcal{L}$ for a spinor field $\psi$ in curved spacetimes \cite{guideForCalculations},
    \begin{equation}
        \mathcal{L} = \frac{1}{2}\sqrt{-g}\bar{\psi}\left(i \gamma^\mu\nabla_\mu -m\right)\psi +\textrm{H.c.},
    \end{equation}
    where $\bar{\psi} = \psi^\dagger \gamma^0$ and $\gamma^\mu = e^\mu_I\gamma^I$ in any coordinate system. When extremized, it yields Dirac's equation as an equation of motion,
    \begin{equation}\label{diracCurved}
        \left(i \slashed \nabla -m\right)\psi = 0.
    \end{equation}
    This is the natural generalization of Eq. \eqref{diracFlat}, now taking into account the connection $\nabla_\mu$. Feynman's convention is still valid, using the \b general \c covariant $\gamma^\mu$ matrices. \b Notice that in order for equation Eq. \eqref{diracCurved} to yield unique solutions given initial data in a spacelike surface, one must assume a regularity condition for the spacetime $\mathcal{M}$. Namely, we must assume $\mathcal{M}$ to be a globally hyperbolic~\cite{Wald1,hawking} spacetime, so that given initial data in a Cauchy surface, it is possible to obtain the solution for the field in any point of spacetime.\c
    
    %Having this it should be possible to make an expansion analogous to what was shown in the section above that would be valid in a general spacetime $\mathcal{M}$. For that we will need to understand clearly what step would be the equivalent of the nonrelativistic approximation made in \eqref{approxFlat}.
    
    The solutions to Dirac's equation provide a relativistic quantum mechanical description of a spin $\frac{1}{2}$ fermion. Indeed, it is possible to obtain a probability density from the following conserved current
    \begin{equation}
        J^\mu = \bar{\psi} \gamma^\mu\psi.
    \end{equation}
    Given a four velocity $u^\mu$, the quantity $u_\mu J^\mu$ can be interpreted as a probability density for the Dirac wavefunction $\psi(x)$. This motivates the definition of the inner product between two spinors $\phi(x)$ and $\psi(x)$ in a spacelike hypersurface \b $\Sigma$ \c with future pointing unit normal $n^\mu$ to be
    \begin{equation}\label{spinInnerProd}
        (\phi,\psi) =  \int_{\Sigma}\dd\Sigma\, \bar{\phi} n_\mu \gamma^\mu \psi,
    \end{equation}
    which is independent of $\Sigma$. \tb{It is then possible to identify the Hilbert space associated with the underlying quantum theory to be the space of spinors defined in a given Cauchy surface, equipped with the inner product from Eq. \eqref{spinInnerProd}.} It is with respect to this inner product that it is possible to apply the standard techniques and interpretations of quantum mechanics. For example, $(\phi,\psi)$ can be understood as the probability amplitude for a state described by the spinor $\psi$ to be measured in the state $\phi$. It is worth \tb{pointing} out that this prescription can be used to describe an electron under the influence of a central charge, describing a hydrogen-like atom and this formulation gives more accurate predictions for the energy levels than the nonrelativistic description \cite{dirac}.
    
\section{Synge's World Function and Fermi Normal Coordinates}\label{secSynge}
    
    \b
    In Section \ref{secSchAsDir} we reduced Dirac's equation to Schr\"odinger's equation in an inertial coordinate system. In order to extend this reduction to curved spacetimes, we consider the generalization of inertial coordinates to arbitrary timelike trajectories: the Fermi normal coordinates (FNC).
    \c
    In this section we review the definition and basic properties of Synge's world function in order to construct the FNC~\b\cite{MisnerFermiNormal,poisson} \c around a timelike trajectory $z(\tau)$. We then present an expansion of the metric and connection to first order in the acceleration of the curve and spacetime curvature in these coordinates. \b These expansions will allow us to compute the spin connection and Dirac's equation in these coordinates in Section \ref{secDirWorldline}.\c
    
    Synge's world function, among other features, allows one to generalize the concept of separation vector \b locally in any \c  curved spacetime \cite{Synge1931}. It is a real scalar function that takes two spacetime points as input and is defined as half the square of their geodesic distance,
    \begin{equation}
        \sigma(x',x) = \frac{1}{2}(u_1-u_0)\int_{u_0}^{u_1} g_{\mu\nu} \dot{\gamma}^\mu(u)\dot{\gamma}^\nu(u)du,
    \end{equation}
    where $\gamma(u)$ is assumed to be the unique geodesic connecting $x$ and $x'$ such that $\gamma(u_0) = x$ and $\gamma(u_1) = x'$, where $u$ is any parameter. Therefore, Synge's function is only defined within a very special region of $\mathcal{M}\times \mathcal{M}$, where the points $(x',x)$ can be connected by a unique geodesic.
    
    Given that Synge's world function depends on two spacetime points, it can be differentiated with respect to each of these. Its total covariant derivative with respect to $x$ is then a $1$-form in $T^*_x\mathcal{M}$, while the covariant derivative with respect to $x'$ is an element of $T^*_{x'}\mathcal{M}$. To handle this, it is important to keep track of the point with respect to which we are differentiating. In order to do this we will use the convention that primed indices refer to differentiation with respect to the first argument and unprimed indices to the second.
    
    As an example, Synge's function in Minkowski spacetime and inertial coordinates can be written simply as
    \begin{equation}
        \sigma(x',x) = \frac{1}{2}(x'-x)^\mu(x'-x)_\mu.
    \end{equation}
    In particular, this means that in flat spacetimes the separation vector between $x'$ and $x$ can be obtained by differentiating $\sigma$ with respect to its arguments,
    \begin{align}
        (x'-x)_\mu = \pdv{\sigma}{x^{\mu'}}, &&&-(x'-x)_\mu = \pdv{\sigma}{x^\mu}.
    \end{align}
    As we will see, an analogue property holds in curved spacetimes.
    
    In general curved spacetimes, the derivatives of Synge's world function can be used to define an analogous concept to the separation vector in flat spacetimes. Indeed, let $\gamma(s)$ be the unique geodesic that connects the point $x$ to the point $x'$, parametrized by arc length. Then, it is possible to show \cite{dirac} that 
    \begin{align}
        \dot{\gamma}_{\alpha'} = \pdv{\sigma}{x^{\alpha'}},  &&& \dot{\gamma}_{\alpha} = -\pdv{\sigma}{x^{\alpha}},
    \end{align}
    where $\dot{\gamma}_{\alpha'}$ is the final tangent vector to the geodesic and $\dot{\gamma}_{\alpha}$ is its initial velocity. This property allows for the generalization of the concept of separation vector between two points to be defined as the initial velocity of the geodesic that connects them.
    
    Synge's world function is useful in many contexts of general relativity and geometry in general. For example, given a point in spacetime, the Riemann normal coordinates around this point can be defined in terms of the derivatives of $\sigma(x',x)$. Another use of  Synge's world function is to define covariant tensor expansions locally around a point, which is a fundamental tool for many perturbative approaches to general relativity \cite{poisson,DixonI,DixonII,DixonIII}.

    Synge's world function can also be used to define the FNC. These are coordinates associated with a given observer that undergoes a timelike curve $z(\tau)$, which we assume to be parametrized by proper time $\tau$. It is a coordinate system $x = (\tau,\bm x)$ such that $\tau$ is the proper time of the curve and $\bm x$ defines coordinates in the rest space associated with $z(\tau)$ for each $\tau$. We define the rest space associated with $z(\tau)$ to be the spacelike hypersurface $\Sigma_\tau$ defined by the events $x$ such that
    \begin{equation}\label{restSpace}
        \nabla_\alpha\sigma(x,z(\tau))u^\alpha(\tau) = 0,
    \end{equation}
    where $u^\alpha(\tau)$ denotes the four-velocity of the curve at time $\tau$. In essence, equation \eqref{restSpace} defines $\Sigma_\tau$ to be the surface reached by every geodesic originating from the point $z(\tau)$ which has its initial tangent vector orthogonal to $u^\alpha(\tau)$.
    
    The FNC can then be defined in terms of an orthonormal frame $(e_I(\tau))$, where we write $I = (0,\mathrm{i})$, in agreement with the previous convention. This frame is defined along $z(\tau)$ such that $e_0(\tau) = u(\tau)$ and the $e_{\mathrm{i}}(\tau)$ are spacelike. This frame is imposed to be Fermi-Walker transported along the curve. That is, once the basis $(e_I(\tau_0))$ is chosen at a given instant of time $\tau_0$ with $e_0(\tau_0) = u(\tau_0)$, we extend it to the whole curve by imposing the following first order differential equation\b
    \begin{equation}\label{FWtransp}
        \frac{D e_I^\mu}{\dd\tau}  + (a^{\mu}u^{\nu}-a^{\nu}u^{\mu})(e_{I})_{\nu} = 0,
    \end{equation}\c
    where $a^\mu$ is the four acceleration of the curve. The equation above imposes that the frame is irrotational from the perspective of the curve (more about this can be found in~\cite{poisson}). In particular, it is easy to see that $u(\tau)$ satisfies the differential equation above for all $\tau$, implying that once $e_0(\tau_0)$ is chosen to be the four-velocity, it will coincide with $u(\tau)$ along the whole curve. To build the FNC, we associate the coordinate $x = (\tau,\bm x)$ to the point $x$ which is reached by the geodesic that starts at $z(\tau)$ with initial velocity $x^{\mathrm{i}}e_{\mathrm{i}}(\tau)$. From this point on, the only coordinate system we will use throughout the manuscript will be the FNC associated with a given curve, so we will stick to the convention that Greek indices are split into its space and time components according to $\mu = (\tau, i)$ and refer to components in the corresponding coordinate basis. Notice that the space indices associated with the FNC are italic, unlike the ones associated with the frame.

    With this construction, the proper distance between a point $x$ with coordinates $x = (\tau,\bm x)$ and the curve will be given by the Euclidean norm of $\bm x$, that is $r=\norm{\bm x} = \sqrt{x_i x^i}$. This gives a physical meaning to this coordinate system, associating $\tau$ to the proper time of the curve and $\bm x$ to the proper distance between an event and the trajectory. In a way, the FNC can be seen as the generalization of the coordinates associated with inertial trajectories in Minkowski spacetimes.%, which have a natural physical meaning. 
    
    Using the tensor expansions from \cite{poisson}, it is possible to expand the metric in Fermi normal coordinates in terms of the proper distance between a point and the curve. The expansion to first order in acceleration and curvature reads
    \begin{equation}\label{expansionFNC}
        \begin{aligned}
            &g_{\tau \tau}=-\left(1+2 a_{{\mathrm{i}}}(\tau)  x^{i}+R_{0 {\mathrm{i}} 0 {\mathrm{j}}}(\tau)  x^{i}  x^{j}\right),\\
            &g_{\tau i}=-\frac{2}{3} R_{0 {\mathrm{j}}{\mathrm{i}}{\mathrm{k}}}(\tau)  x^{j}  x^{k},\\
            &g_{ij}=\delta_{{\mathrm{i}}{\mathrm{j}}}-\frac{1}{3} R_{{\mathrm{i}}{\mathrm{k}}{\mathrm{j}}{\mathrm{l}}}(\tau)  x^{k}  x^{l},
        \end{aligned}
    \end{equation}
    where $R_{IJKL}(\tau)$ denotes the Riemann curvature tensor evaluated along the curve $z(\tau)$ and $a_I(\tau)$ denotes the proper acceleration of the worldline. Notice that in the expansion above the curvature tensor and acceleration components are evaluated in the frame, while the FNC components naturally have italic indices. For brevity we will not explicitly write the $\tau$ dependence in the components of the curvature tensor and acceleration. \b The expansion in Eq. \eqref{expansionFNC} is well defined in the whole normal convex neighbourhood of the curve $z(\tau)$, where Synge's world function is defined~\cite{poisson}. \c The computation of the volume form in Fermi normal coordinates can be found in \cite{us} and reads
    \begin{align}\label{sqrtG}
        \sqrt{-g} = 1 + a_{\mathrm{i}}x^i +\left(\frac{1}{3}R_{0 {\mathrm{i}} 0 {\mathrm{j}}}-\frac{1}{6}R_{{\mathrm{i}}{\mathrm{j}}}\right) x^ix^j.
    \end{align}
    
    %Now, our goal is to pick a curve $z(\tau)$ associated with an observer and consider an expansion of the sort of the one of Equation \eqref{approxFlat} so that we can obtain an equation of motion for an effective wave function. To do that, we first notice that the equation of motion for such wavefunction has to be, to zeroth order in the expansion and on curvature, a second order differential equation. This means that we will have to expand the metric to fourth order in order to calculate the Christoffel symbols and their derivatives. To perform such an expansion we follow the approach of Poison from \cite{poisson}.
    
    From the metric we can compute the Christoffel symbols in these coordinates to first order in curvature and acceleration. To do so, and throughout the manuscript we will work under the assumption that derivatives of curvature and acceleration are of subleading order. Under this assumption, the Christoffel symbols are given by 
    \begin{align}
        \Gamma_{i j}^{\tau}&=\frac{1}{3}\left(R_{0 {\mathrm{i}}{\mathrm{j}}{\mathrm{m}}}+R_{0 {\mathrm{j}}{\mathrm{i}}{\mathrm{m}}}\right) x^{m},\label{gamma1}\\
        \Gamma_{\tau i}^{\tau}&=a_{\mathrm{i}} +R_{0 {\mathrm{i}} 0 {\mathrm{m}}} x^{m}, \label{gamma2}\\
        \Gamma_{\tau\tau}^{\tau}&=0,\label{gamma3}\\
        \Gamma_{j k}^{i}&=\frac{1}{3}\left(R\indices{_{\mathrm{j}}^{\mathrm{i}}_{\mathrm{k}\mathrm{m}}}+R\indices{_{\mathrm{k}}^{\mathrm{i}}_{\mathrm{j}\mathrm{m}}}\right) x^{m}, \label{gamma4}\\
        \Gamma_{\tau j}^{i}&=R\indices{_{0 {\mathrm{m}} {\mathrm{j}}}^{\mathrm{i}}} x^{m},\label{gamma5}\\ \Gamma_{\tau\tau}^{i}&=a^{\mathrm{i}}+R\indices{_{0}^{\mathrm{i}}_{ 0 {\mathrm{m}}}} x^{m}\label{gamma6}.
    \end{align}
    With this, we have the means to perform tensor calculations in Fermi normal coordinates up to first order in acceleration and curvature.
    
    Although we have built the Fermi normal coordinates, in order to work with spinors we must also have a locally defined frame such that we can apply the techniques described in Section \ref{secSpniCur}. In this sense, the frame $e_I$ is still only defined along the curve and not in a neighbourhood of it. To extend the frame to a  local region around the curve $z(\tau)$, we parallel transport the vectors $e_I(\tau)$ along the spacelike geodesics used to build the FNC. This means that we can now refer to the frame $e_I(x)$, bearing a spacetime dependence and localized around the curve. It is important to remark that although this frame is parallel transported, $e_0$ is in general not normal to the surfaces and neither are the vectors $e_{\mathrm{i}}$ tangent to $\Sigma_\tau$.
    
    It is then possible to express the components of the frame vectors in the Fermi normal coordinate basis. We thus obtain the following expressions for the frame to first order in curvature and acceleration,
    \begin{align}
        e_{0}^{\mu}&=\delta_{\tau}^{\mu}(1-a_{\mathrm{i}} x^i)+\frac{1}{2} R\indices{^{\mu}_{{\mathrm{l}} 0 {\mathrm{m}}}} x^{l} x^{m} \\
        e_{{\mathrm{i}}}^{\mu}&=\delta_{i}^{\mu}+\frac{1}{6} R\indices{^{\mu}_{{\mathrm{l}}{\mathrm{i}}{\mathrm{m}}}} x^{l} x^{m}.
    \end{align} 
    \b We remind the reader that $I = (0,\text{i})$ refers to the components in the orthonormal frame and $\mu = (\tau,i)$ to the components in the Fermi normal coordinates. \c The dual frame's components can be calculated by inverting the matrix of components above and reads
    \begin{align}
        e_{\tau}^{I}&=\delta_{0}^{I}(1+a_{\mathrm{i}} x^i)-\frac{1}{2} R\indices{^{I}_{{\mathrm{l}} 0 {\mathrm{m}}}} x^{l} x^{m} \\
        e_{i}^{I}&=\delta_{{\mathrm{i}}}^{I}-\frac{1}{6} R\indices{^{I}_{{\mathrm{l}}{\mathrm{i}}{\mathrm{m}}}} x^{l} x^{m}.
    \end{align} 
    Finally, to be able to calculate the spin connection one requires the connection on the orthonormal frame. In Appendix \ref{appFrame} we compute the components of the connection $1$-form associated with the frame to first order in curvature and acceleration,   
     \begin{align}
        \omega\indices{_{\mu}^0_0} &=  0,\label{omega1}\\
        \omega\indices{_\tau^{{\mathrm{i}}}_0} &=  a^{\mathrm{i}}+R\indices{_{0}^{\mathrm{i}}_ {0{\mathrm{m}}}} x^{m},\label{omega2}\\
        \omega\indices{_i^{{\mathrm{j}}}_0}&=   \frac{1}{2}R\indices{_{0}^{\mathrm{j}}_{\mathrm{im}}}x^m,\label{omega3}\\
        \omega\indices{_{\tau}^{{\mathrm{i}}}_{{\mathrm{j}}}} &= R\indices{_{0\mathrm{mj}}^{\textrm{i}}}x^m,\label{omega4}\\
        \omega\indices{_k^{{\mathrm{j}}}_{{\mathrm{i}}}} &= \frac{1}{2}R\indices{_{\mathrm{i}}^{\textrm{j}}_{\textrm{km}}}x^{m}.\label{omega5}
    \end{align}
    With these we have all the tools needed to compute the effect of curvature and acceleration on the description of a Dirac spinor. In the next section we will compute the spin connection and the correction terms that arise in the Hamiltonian of a spinor system due to curvature and acceleration.
    
\section{Dirac's Equation around a Worldline}\label{secDirWorldline}
    
    The goal of this section is to apply the results of Section \ref{secSynge} and write Dirac's equation for a spinor in curved spacetimes in Fermi normal coordinates around a given worldline $z(\tau)$. We will then expand the equation to first order in acceleration and curvature in order to obtain the system's Hamiltonian. This will allow us to obtain the effects of curvature and acceleration on the dynamics of the system. 
    
    We first write Eq. \eqref{diracCurved} in Fermi normal coordinates. In terms of the spin connection, Dirac's equation can be written as
    \begin{align}
        \b 0=&(i\slashed \nabla - m)\psi= (i \gamma^\tau \nabla_\tau + i \gamma^i\nabla_i -m)\psi \\
        &=  (i \gamma^\tau \partial_\tau + i \gamma^\tau \Gamma_\tau + i \gamma^i\partial_i+i \gamma^i \Gamma_i -m)\psi,\c
    \end{align}
    where \b the components of the spin connection operators \c $\Gamma_\mu$ are given by Eq. \eqref{ChristSpin}. In order to get the Hamiltonian, we factor the time evolution term $i \partial_\tau \psi$ to the left hand side and obtain
    \begin{align}
        i \partial_\tau \psi =-(\gamma^\tau)^{-1}(i \gamma^i\partial_i+i \gamma^i \Gamma_i -m)\psi-i \Gamma_\tau \psi.
    \end{align}
    The terms on the right hand side are then identified as the Hamiltonian operator $H$, that generates time translation with respect to the proper time of the curve $\tau$. $H$ can be seen as generating time evolution between the rest space hypersurfaces associated with the trajectory. The Hamiltonian is then given by
    \begin{align}
        H = -(\gamma^\tau)^{-1}(i \gamma^i\partial_i+i \gamma^i \Gamma_i -m)\psi-i \Gamma_\tau .
    \end{align}
    
    In order to work with the expression above, it is useful to rewrite the inverse of $\gamma^\tau$ in terms of other known operators. For that, we use the fact that
    \begin{equation}
        (\gamma^\tau)^{-1} =- (g^{\tau\tau})^{-1}\gamma^\tau,
    \end{equation}
    which can be seen %multiplying
    %\begin{equation}
         %(\gamma^{\tau})^{-1} \gamma^\tau = (g^{\tau \tau})^{-1} \gamma^\tau \gamma^\tau = (g^{\tau \tau})^{-1} g^{\tau \tau}  = 1
    %\end{equation}
    from \mbox{$\{\gamma^\tau, \gamma^\tau\} = -2 g^{\tau\tau}$}. Therefore, the Hamiltonian can be written as
    \begin{align}\label{HDiracFNC}
        H = (g^{\tau\tau})^{-1}\gamma^\tau(i \gamma^i\partial_i+i \gamma^i \Gamma_i -m)\psi-i \Gamma_\tau.
    \end{align}
    
    The spin connection $\Gamma_\mu$ can be expanded to first order in curvature and acceleration using Eq. \eqref{ChristSpin}. In Appendix \ref{appDirH} we compute the expansions in Fermi normal coordinates and obtain
    \begin{align}
        \Gamma_\tau &= -\frac{1}{2}(a_{\mathrm{i}}+R_{0{\mathrm{i}}0{\mathrm{m}}}x^m) \gamma^{{\mathrm{i}}}\gamma^0 - \frac{1}{4}R_{0 {\mathrm{mij}}}x^m\gamma^{{\mathrm{i}}}\gamma^{{\mathrm{j}}},\\
        \Gamma_{k} &= -\frac{1}{4}R_{0{\mathrm{ikm}}}x^m\gamma^{{\mathrm{i}}}\gamma^0 - \frac{1}{8} R_{{\mathrm{jikm}}} x^m \gamma^{{\mathrm{i}}}\gamma^{{\mathrm{j}}}.
    \end{align}
    Plugging the results above in Eq. \eqref{HDiracFNC} and expanding to first order in acceleration and curvature yields the following expression for the Hamiltonian
    \begin{widetext}
    \begin{align}\label{HDiracExp}
        H &=  -\gamma^0(i \gamma^{{\mathrm{i}}}\partial_i -m)  -  a_{\mathrm{i}} x^i\gamma^0 \gamma^{{\mathrm{i}}}i\partial_i - \frac{1}{2}R_{{\mathrm{k}}0{\mathrm{m}}0}x^kx^m\gamma^0i \gamma^{{\mathrm{i}}}\partial_i +m a_{\mathrm{i}}x^i\gamma^0+\frac{m}{2}R_{{\mathrm{k}}0{\mathrm{m}}0}x^kx^m\gamma^0  +\frac{1}{6} R\indices{_{0{\mathrm{ljm}}}} x^{l} x^{m}\gamma^{{\mathrm{j}}}\gamma^{{\mathrm{i}}}i\partial_i\nonumber \\
        &-\frac{i}{2}R\indices{^{\mathrm{i}}_{{\mathrm{l}}0{\mathrm{m}}}}x^lx^m\partial_i-\frac{m}{6} R\indices{_{0{\mathrm{ljm}}}} x^{l}x^m\gamma^{{\mathrm{j}}} -\frac{i}{6}R\indices{^{\mathrm{j}}_{{\mathrm{lim}}}}x^lx^m\gamma^0\gamma^{{\mathrm{i}}}\partial_j-\frac{i}{4}\left(2a_{\mathrm{i}} + R_{0{\mathrm{i}}0{\mathrm{m}}}x^m-R_{{\mathrm{im}}}x^m\right) \gamma^0 \gamma^{{\mathrm{i}}}-\frac{i}{4}R_{0{\mathrm{imj}}}x^m\gamma^{{\mathrm{i}}}\gamma^{{\mathrm{j}}}.
    \end{align}
    \end{widetext}
    It is important to remark that all the curvature terms are evaluated along the curve $z(\tau)$, so that the $R_{IJKL}$ and the $a_\mathrm{i}$ depend only on the time parameter $\tau$. The result above agrees with Parker's result from \b \cite{ParkerRevLett,parker} \c if one sets the acceleration to zero. 

    It is also important to comment on the Hermiticity of this Hamiltonian with respect to the conserved inner product from Eq. \eqref{spinInnerProd}. The conserved inner product can be compared to the ``flat'' one by means of the expansion in terms of curvature and acceleration. Indeed, the conserved inner product is given by
    \begin{align}\label{inprodSpinor}
        (\phi,\psi) = \int_{\Sigma}\dd\Sigma_\mu \bar{\phi}(x)  \gamma^\mu(x) \psi(x),
    \end{align}
    which can then be expanded in the space slices associated with the curve $z(\tau)$. To perform the integral of Eq. \eqref{inprodSpinor} one must notice that $e_0$ is not necessarily orthogonal to the surface at every point. That is, it is not true that $n_\mu \gamma^\mu = \gamma^0$ in general. We can then use a $3+1$ metric decomposition associated with the FNC. The spinor inner product takes the form presented in \b \cite{ParkerRevLett,parker}\c, which reads
    \begin{align}\label{inprodSpinorParker}
        (\phi,\psi) = \int_{\Sigma}\dd^3 x \:\sqrt{-g}  \bar{\phi}(x)  \gamma^\tau(x) \psi(x).
    \end{align}
    The equivalence between Eqs. \eqref{inprodSpinor} and \eqref{inprodSpinorParker} can be found in \cite{guideForCalculations}.
    
    Using the fact that $\gamma^\tau = e^\tau_I \gamma^I$ and the expansion for $\sqrt{-g}$ from Eq. \eqref{sqrtG}, one finds that the spinor inner product can be written as
    \begin{align}\label{curInnerFlat}
        (\phi,\psi) = (\phi,\psi)_0 + (\phi,C \psi)_0,
    \end{align}
    where $C$ is defined by
    \begin{align}
        C &= \sqrt{-g}\gamma^0\gamma^\tau - \openone \\&= -\frac{1}{6}\left((R_{0\mathrm{i}0\mathrm{j}}+R_{\mathrm{i}\mathrm{j}})x^ix^j+R_{0\mathrm{ikj}}x^ix^j\gamma^0\gamma^{{\mathrm{k}}}\right).
    \end{align}
    Notice that the inner product is independent of the acceleration of the worldline, it only depends on the curvature of spacetime. We define $ (\phi,\psi)_0$ in analogy with the inner product for flat spacetimes, but in terms of the FNC in the constant $\tau$ surfaces,
    \begin{equation}
         (\phi,\psi)_0 = \int \dd^3 x \phi^\dagger(x) \psi(x).
    \end{equation}    
    We have checked that the acceleration terms in the Hamiltonian from Eq. \eqref{HDiracExp} are symmetric with respect to the covariant inner product to the order we are considering. The remaining terms have been checked in \b \cite{ParkerRevLett,parker}\c. The Hermiticity condition is important so that the Hamiltonian generates a $unitary$ time evolution. 
    
\section{Schr\"odinger Equation in Curved Spacetimes}\label{secSchInCur}

    In this section we reduce Dirac's Hamiltonian for a spinor \tb{particle} to an operator that acts on complex wavefunctions defined in spacelike surfaces $\Sigma_\tau$. We take a similar approach to the one described in Section \ref{secSchAsDir}, but we start from the Hamiltonian in Eq. \eqref{HDiracExp} and only consider first order terms in acceleration and curvature. However, there are two problems that have to be considered. First, one must choose  a worldline associated with the quantum system such that the expansion \eqref{expansionFNC} is valid. Second, to provide a consistent quantum theory for the wavefunctions defined in the rest spaces, one needs to define \tb{a Hilbert space and} an inner product such that the Hamiltonian operator is Hermitian. 
    
    In order to approach these problems, we start by pointing out what was done in Section \ref{secSchAsDir}, where we have seen that for inertial trajectories in \b Minkowski spacetime\c, one could describe \tb{a quantum particle} by a nonrelativistic wavefunction defined in the constant time slices. In terms of coordinates $(t,\bm x)$ such that $\bm x$ provide coordinates in the space slices, one could then do standard quantum mechanics in the \tb{Hilbert} space of square integrable functions in each slice. The natural inner product under these assumptions is then given by
    \begin{equation}\label{flatty}
        (\phi,\psi) = \int \dd^3 x \phi^*(x)\psi(x).
    \end{equation}
    Of course this procedure is not covariant, or valid in other reference frames, rather, it would only provide an approximate description of a localized quantum system in its own reference frame. \b Notice that it is not trivial to provide a similar formalism in the context of curved spacetimes, where not even the choice of time slice is clear. In fact, to the authors knowledge, \c the generalization of this wavefunction approximation to the general case of arbitrary trajectories in curved spacetimes has not yet been studied.
    
    To generalize this approach to curved spacetimes, we perform a similar approximation around a trajectory $z(\tau)$. In order for the expansion of Eq. \eqref{expansionFNC} to be valid, one must assume that the average extension of the system is small compared to the curvature radius. The natural worldline to pick is then the trajectory of the center of mass of the system. To define the center of mass worldline one can use Dixon's definition \cite{DixonI,DixonII,DixonIII}, which only requires the stress-energy tensor of the system. The trajectory $z(\tau)$ then defines a natural notion of rest space where the wavefunctions can be defined. That is, in Fermi normal coordinates, the system can be described by a complex wavefunction $\psi(\tau,\bm x)$. For every value of $\tau$, $\psi(\tau,\bm x)$ can be seen as a function defined in the surface $\Sigma_\tau$. Therefore, the Hilbert space used to define our \tb{one-particle} quantum theory will be a space of square integrable functions in $\Sigma_\tau$ with respect to a given measure for each $\tau$. The major problem with this idea is that in principle there is no unambiguous way to choose the integration measure that defines the inner product. Nevertheless, there are two main candidates that we are going to call respectively the flat and the curved inner products
    \begin{align}
        (\phi,\psi)_0 &= \int \dd^3  x \phi^*( x)\psi( x)\label{innerProdSigmaFlat},\\
        (\phi,\psi) &= \int \dd^3  x \,\sqrt{g_\Sigma} \phi^*( x)\psi( x),\label{innerProdSigma}
    \end{align}
    where $\sqrt{g_\Sigma}\dd^3  x = \dd\Sigma $ is the volume element of $\Sigma_\tau$.
    
    While the flat inner product could be seen as natural due to its similarity to the one from Eq. \eqref{flatty} and the fact that it neglects the gravitational effects, the curved inner product is more natural precisely due to the fact that it accommodates for the intrinsic geometry of the surfaces. 
    %in the sense that derivatives with respect to the surface's intrinsic geometry would allow for integration by parts. On the same line of reasoning, the spatial momentum could be associated with an intrinsic derivative on the space slices in a way that it would be covariant. 
    Nevertheless, the flat inner product is not only easier to work with, but can also be compared to the curved one by means of the curvature and acceleration expansions we have been considering so far. Indeed, the difference between these inner products will be given by the factor of $\sqrt{g_{\Sigma}}$, whose expansion can be found to be, see e.g. \cite{us},
    \begin{align}
        \sqrt{{g}_\Sigma} &= 1 - \frac{1}{6}\left(R_{1\mathrm{i}1\mathrm{j}} + R_{2\mathrm{i}2\mathrm{j}} + R_{3\mathrm{i}3\mathrm{j}} \right)x^ix^j\\
        &= 1-\frac{1}{6}\left(R_{0\mathrm{i}0\mathrm{j}} + R_{\mathrm{i}\mathrm{j}}\right)x^ix^j.
    \end{align}
    Although we will make use of both the flat and curved inner products, the one that should be used for the computation of expected values and transition probabilities is the one that makes the Hamiltonian Hermitian. It is important to remark that the choice of inner product is not a cosmetic problem, for it might yield radically different predictions for transition probabilities and expected values.
    %Of course, choosing between the flat and curved spacetime options is a matter of choice. Might very well be the case that neither of them make the Hamiltonian Hermitian. Who knows? Therefore, let's not make any choice at all. Instead, let us grab that beautiful covariantly conserved inner product for spinors and reduce it to a inner product for complex wavefunctions. 
    
    There is also the possibility that neither the flat nor the curved inner products make the Hamiltonian symmetric. Indeed, the natural inner product to be used should come from the fully covariant spinor theory. We then reduce the spinor inner product to an inner product for wavefunctions by tracing over the spin degrees of freedom. Although this does not guarantee that the Hamiltonian will be Hermitian, it is consistent with our general approach. %The approach we take is to trace over the spin degrees of freedom of the spinor inner product  to   in this work, where we use the calculations from the previous sections to reduce the formalism for spinors in curved spacetimes to an approximate formalism for a complex wavefunction. 
    In order to reduce the spinor inner product, we look at Eq. \eqref{curInnerFlat}, where it is expanded in terms of the flat inner product and a correction term $C$. If one traces over the spin components of $C$, one finds that the inner product for a complex wavefunction should then be, to first order in curvature and acceleration,
    \begin{equation}\label{innerProdSigmaExp}
        (\phi,\psi) = \int \dd^3 x \phi^*(x)\left(1-\frac{1}{6}\left(R_{0\mathrm{i}0\mathrm{j}} + R_{\mathrm{i}\mathrm{j}}\right)x^ix^j\right)\psi(x).
    \end{equation}
    Notice that this is exactly what one obtains if one expands the measure on the surfaces $\sqrt{g_{\Sigma}}$ in terms of curvature. This implies that, in principle, the covariant inner product from Eq. \eqref{innerProdSigma} is equivalent to tracing out the spin degrees of freedom of the covariantly conserved inner product from Eq. \eqref{inprodSpinor}. Indeed, we will also show that the Hamiltonian obtained for the wavefunction description will be Hermitian with respect to this inner product.
        
    We now apply the analogue of the approximation from Eq. \eqref{approxFlat} to this general setup. That is, we solve for the $B$ component of the spinor in terms of the $A$ component and plug this result back into the Hamiltonian. It is possible to split the time evolution equation associated with the Hamiltonian from Eq. \eqref{HDiracExp} in terms of its $A$ and $B$ components,
    \begin{align}
        i\partial_\tau \psi_A = H_{AA}\psi_A + H_{AB}\psi_B,\label{eqA}\\
        i\partial_\tau \psi_B = H_{BA}\psi_A + H_{BB}\psi_B.\label{eqB}
    \end{align}
    We then solve Eq. \eqref{eqB} for $\psi_B$ in terms of $\psi_A$, and obtain
    \begin{equation}\label{generalIdea}
        \psi_B = (i\partial_\tau - H_{BB})^{-1}H_{BA}\psi_A\equiv D_B \psi_A,
    \end{equation}
    where $D_B$ is a matrix valued differential operator. Although in principle the solution for $D_B$ can be done exactly, a perturbative approach is able to yield Schr\"odinger's description from Dirac's. For this non-relativistic limit, we consider a system whose energy is much lower than its rest mass. Similar to the case in Minkowski spacetime, we expand the inverse operator in Eq. \eqref{generalIdea} as a power series in $m^{-1}$. In order to do so, we must keep track of two approximations. First, we only consider terms of first order in acceleration and curvature. Second, we neglect terms of the order $m^{-2}$ or higher. This is justified, because the rest mass is assumed to be much larger than acceleration, curvature or the non-relativistic energy of the system. A more detailed discussion about these approximations can be found in Appendix \ref{appSchCur}.
    
    The next step is to plug the result of Eq. \eqref{generalIdea} back into Eq. \eqref{eqA} in order to obtain a differential equation only for the $A$ component of the spinor, $\psi_A$. We obtain
    \begin{equation}
        i \partial_\tau \psi_A = (H_{AA} + H_{AB}D_B)\psi_A \equiv \mathcal{H}_A\psi_A.
    \end{equation}
    $\mathcal{H}_A$ can then be regarded as the effective Hamiltonian for the two-component wavefunction $\psi_A$. In order to obtain a single-valued complex wavefunction description, we then trace over the spin degrees of freedom on the Hamiltonian level. Thus, the Hamiltonian $\mathcal{H}$ compatible with the description of a Schr\"odinger wavefunction $\psi(x)$ in the reference frame associated with an observer undergoing a trajectory $z(\tau)$ in curved spacetimes can be written as
    \begin{equation}
        \mathcal{H} = \tr\mathcal{H}_A.
    \end{equation}
    The calculations regarding the curvature and acceleration expansions are done in detail in Appendix \ref{appSchCur}.   
    
    We obtain the following Hamiltonian operator for a complex wavefunction $\psi(x)$,
    \begin{widetext}
    \begin{align}
        \mathcal{H} =& \:m - \frac{1}{2m}\partial^i\partial_i +m a_{\textrm{j}}x^j+\frac{m}{2}R_{\textrm{k}0\textrm{m}0}x^kx^m  -\frac{2}{3} R\indices{_{0\textrm{lim}}} x^{l} x^{m}i\partial^i +\frac{i}{3}R_{0\textrm{k}}x^k\nonumber\\
        &-\frac{3}{4m}\left( a_\textrm{j} x^j  + \frac{1}{2}R_{\textrm{k}0\textrm{m}0}x^kx^m\right) \partial_i \partial^i-\frac{3}{4m}\left(a_{\mathrm{i}} +R_{0\textrm{i}0\textrm{k}}x^k \right)\partial^i \label{YES}\\
        &-\frac{1}{6m}R\indices{_{\textrm{jlim}}}x^lx^m \partial^i\partial^j-\frac{1}{3m}\delta^{jm}R_{\textrm{jlim}}x^l\partial^i +\frac{R}{8m}.\nonumber
    \end{align}
    \end{widetext}
    Eq. \eqref{YES} is the major result of this work. It prescribes the Hamiltonian that generates the time evolution of a wavefunction with respect to its proper time. Notice that the rest mass of the system $m$ contributes with a term proportional to the identity, and therefore only with a global phase. We also obtain a correction term to the rest energy, proportional to the Ricci scalar $R$. There are also two terms proportional to the mass of the system, which are expected to contribute the most: The $ma_{\textrm{i}} x^i$ term is associated with the energy contribution of the external force accelerating the system, while the term proportional to curvature is associated with the tidal forces acting on the system due to spacetime curvature.

    \fix{It is worth mentioning that the protocol outlined here ignores the spin degrees of freedom of the spinors. This is different from taking the partial trace over the spin degrees of freedom that could be done in quantum mechanics. That is, we do not consider a full unitary time evolution that takes the spin into account, and later partial trace over these degrees of freedom, which would result in a mixed state. Instead, we obtain the dynamics \emph{only} for the wavefunction of the system, by also partial tracing the Hamiltonian that generates time evolution. This is the standard procedure employed in order to obtain a wavefunction description from a spinor field.}
    
    The coupling of the system to electromagnetism can be done by means of minimal coupling. That is, if one wishes to consider the interaction of the system with an external electromagnetic field $A_\mu$, it is enough to shift
    \begin{equation}
        \partial_\mu \longmapsto D_\mu = \partial_\mu - i q A_\mu. 
    \end{equation}
    The reason this can be done at this stage is detailed in Appendix \ref{appSchCur}, where we perform the calculations considering the electromagnetic field from the beginning.
    
    In Appendix \ref{appHermitian} we show that the Hamiltonian $\mathcal{H}$ is Hermitian with respect to the inner product of Eq. \eqref{innerProdSigmaExp} up to first order in acceleration and curvature. This allows us to associate $\mathcal{H}$ to the energy of the system and ensures that the time evolution generated by this Hamiltonian will be unitary, thus defining a consistent quantum mechanical formulation. The Hilbert space with respect to which this formulation would be done is that of square-integrable functions in the rest spaces $\Sigma_\tau$ for each $\tau$. The final remark is that in general the measure that defines the inner product may be time dependent so that the inner product between two wavefunctions is preserved.
    
    Other observables of the theory also have to be Hermitian with respect to the inner product defined. In particular, we must find self-adjoint operators that are associated with the position and momentum of the system. The position operator $\hat{x}^i$ can always be associated with the product of wavefunctions by the space Fermi normal coordinates. It is indeed a self-adjoint operator with the appropriate physical interpretation. The momentum, however, is not as straightforward. Indeed, the operator $-i \partial_i$ is in general not Hermitian with respect to the inner product of Eq. \eqref{innerProdSigma}. Then, in analogy to what is usually done for quantum mechanics in curved space \cite{Gneiting_2013}, the operator that should be used as the momentum of the system is given by
    \begin{equation}\label{p}
        \hat{p}_i\psi = \frac{-i\:\:\:}{(g_\Sigma)^{\frac{1}{4}}}\pdv{}{x^i}\left((g_{\Sigma})^{\frac{1}{4}}\psi\right).
    \end{equation}
    Indeed, it is the case that this operator satisfies
    \begin{equation}    
       (\phi,\hat{p}_i\psi)= \int \dd\Sigma \phi^*\hat{p}_i\psi = \int \dd\Sigma (\hat{p}_i\phi)^*\psi = (\hat{p}_i\phi,\psi),
    \end{equation}
    showing that it is Hermitian. Not only that, but it also satisfies the canonical commutation relations with the position operator,
    \begin{align}
        \comm{\hat{x}^i}{\hat{p}_j} = i\delta^i_j.
    \end{align}
    Therefore, we argue that the operator defined in Eq. \eqref{p} is the adequate definition of the momentum operator for this system. To first order in curvature, it can then be written as
    \begin{equation}
        \hat{p}_i = -i \partial_i +\frac{i}{6}\left(R_{0 \textrm{i}0\textrm{j}} + R_{\textrm{ij}}\right)x^j.
    \end{equation}
    
    With this we have presented a consistent quantum mechanical theory defined locally around a general worldline $z(\tau)$ in any curved spacetime. In this description, time evolution happens with respect to the proper time of the curve and the wavefunctions can be thought of as complex functions defined in the rest spaces associated with the trajectory. In the next section we will study the coupling with electromagnetism and the hydrogen atom in arbitrary trajectories in curved spacetimes. We also look at the example of a fermion in AdS spacetime. 
    \vspace{5pt}
    
\section{Coupling with electromagnetism and explicit examples}\label{secAtomGravity}

    In this section we provide two examples to the wavefunction formulation presented in Section \ref{secSchInCur} and make general considerations regarding the coupling of these systems with an external electromagnetic field. First we expand Maxwell's equations using the expansions of Section \ref{secSynge}. Then we specialize this discussion to the case of a static atom in Schwarzschild spacetime. The second example is that of a fermion in AdS spacetime, where the negative curvature can act to create a quadratic potential that localizes the system. Overall, we evaluate the typical orders of magnitude of the terms in Eq. \eqref{YES} in specific examples.
    
\subsection{Corrections to the electromagnetic coupling due to acceleration and curvature}

    From Eq. \eqref{YES}, one can compute the Hamiltonian associated with the interaction of a fermionic system of charge $q$ with an external electromagnetic field by shifting the partial derivatives according to $\partial_\mu \longmapsto \partial_\mu - i q A_\mu$. This procedure yields

    \begin{align}\label{HEM0}
        \mathcal{H}_{EM} =& \! -qA_\tau\!+\!\frac{1}{2m}(i q \partial_i A^i\! +\!2iq A_i\partial^i\! +\!q^2 A_i A^i)\! +\! \mathcal{H}^I_{EM},
    \end{align}
    where $\mathcal{H}_{EM}^I$ is the Hamiltonian that contains the first order in curvature and acceleration corrections to the coupling. It is explicitly given by
    \begin{widetext}
        \begin{align}\label{HEM}
        \mathcal{H}_{EM}^I =& \frac{1}{2m}\left(\frac{3}{2} a_\textrm{k} x^k\delta_{ij}  + \frac{3}{4}R_{\textrm{k}0\textrm{m}0}x^kx^m\delta_{ij}+\frac{1}{3}R\indices{_{\textrm{jlim}}}x^lx^m\right)(i q \partial^i A^j +2iq A^i\partial^j +q^2 A^i A^j)\\
        &\quad\quad\quad\quad\quad+\frac{iq}{2m}\left(\frac{3}{2}a_{\textrm{i}} +\frac{3}{2}R_{0\textrm{i}0\textrm{k}}x^k +\frac{2}{3}\delta^{jm}R_{\textrm{jlim}}x^l\right)A^i-\frac{2}{3} qR\indices{_{0\textrm{lim}}} x^{l} x^{m}A^i.\nonumber
    \end{align}
    \end{widetext}
    Notice that the first term in Eq. \eqref{HEM0} does not come from the Hamiltonian \eqref{YES}, instead, it comes from the $i \partial_\tau$ operator in Schr\"odinger's equation. Another remark is that the acceleration/curvature correction terms only involve the vector potential $A_i$, and not $A_\tau$. The reason for this is that in the expansion that takes Dirac's equation to Schr\"odinger's equation does not change the time derivatives, only the spatial ones. %\textcolor{black}{Check the signs of the charges everywhere!}
    
    An important property of Eq. \eqref{HEM} is that it is not gauge invariant. This is a standard feature of Hamiltonians for quantum mechanical systems coupled to electromagnetism. In fact, under gauge transformations $A_\mu \longmapsto A_\mu + \partial_\mu \chi$ for a scalar function $\chi$ the wavefunctions transform according to
    \begin{equation}\label{erickson2}
        \psi(x)\longmapsto e^{-iq\chi(x)}\psi(x).
    \end{equation}
    The Hamiltonian $\mathcal{H}_{EM}$ must then transform according to
    \begin{equation}\label{erickson1}
        \mathcal{H}_{EM}(A_{\textrm{i}}) \longmapsto \mathcal{H}_{EM}(A_{i}+\partial_{i}\chi) -q \partial_\tau \chi
    \end{equation}
    in order to make the equations of motion gauge independent. Equations \eqref{erickson1} and \eqref{erickson2} ensure that the expected value of the energy in any state is a gauge independent quantity.
    
    When considering an external electromagnetic field, it is crucial to compute Maxwell's equations around the system's center of mass worldline. This will be applicable in cases where the sources of Maxwell's theory are comoving with the system or are sufficiently close to it. 
    
    In a general spacetime, Maxwell's equations for the electromagnetic potential can be written as
    \begin{equation}\label{maxwell}
        \nabla^\mu F_{\mu\nu} = -4\pi j_\nu,
    \end{equation}
    where $F_{\mu\nu} = \partial_\mu A_\nu - \partial_\nu A_\mu$ is the electromagnetic tensor. In order to compute the corrections relative to the flat spacetime Maxwell's equations, we assume that the time derivatives of $A_\mu$ can be neglected. This is compatible with our assumption that one can neglect the time derivatives of acceleration and curvature. We will also work in the Coulomb gauge associated with the worldline. That is, we impose the gauge condition $\nabla_i A^i = 0$. Once the expansion of Sec. \eqref{secSynge} is employed, we can write
    \begin{align}
        &\nabla^\mu\nabla_\mu A_\nu =\partial_i\partial^i A_\nu\!+\! a_{\mathrm{i}}\partial^i A_\nu\!-\! \partial^i\Gamma^\alpha_{i\nu}A_\alpha\!-\! 2\Gamma^\alpha_{i\nu}\partial^iA_\alpha\\& \:\:\:\:+\!\left(\!\frac{1}{3}\delta^{kj}R\indices{_{\mathrm{k i j m}}}x^m\!-\!R_{ {\mathrm{i}}  {\mathrm{m}}} x^{m}\!\right)\!\partial^i A_\nu\! +\!\frac{1}{3} R_{{\mathrm{i}}{\mathrm{k}}{\mathrm{j}}{\mathrm{l}}}  x^{k}  x^{l}\partial^i\partial^j A_\nu.\nonumber
    \end{align}
    Plugging in the expressions for the Christoffel symbols of Eqs. \eqref{gamma1},\eqref{gamma2},\eqref{gamma3},\eqref{gamma4},\eqref{gamma5} and \eqref{gamma6}, we obtain the following equation of motion for the time and space components, respectively,
    %\textcolor{white}{If you find this text here, beware. It is simply here because LaTeX is too smart and cannot be tricked in other ways.}
    \begin{widetext}
    \begin{align}
        &\partial_i\partial^i A_\tau +\frac{1}{3} R_{{\mathrm{i}}{\mathrm{k}}{\mathrm{j}}{\mathrm{l}}}  x^{k}  x^{l}\partial^i\partial^j A_\tau-a_{\mathrm{i}} \partial^i A_\tau +2R_{\mathrm{0mji}}x^m\partial^iA^j
        -\left( R_{\mathrm{0i0m}}+\frac{2}{3}\delta^{kj}R_{\mathrm{kijm}}\right)x^m\partial^iA_\tau = - 4\pi j_\tau,\label{Atime}\\
        & \partial_i\partial^i A_l +\frac{1}{3} R_{{\mathrm{i}}{\mathrm{k}}{\mathrm{j}}{\mathrm{m}}}  x^{k}  x^{m}\partial^i\partial^j A_l+\frac{2}{3}(\b R\indices{^\mu_{\mathrm{ilm}}}+R\indices{^\mu_{\mathrm{lim}}}\c )x^m\partial^iA_\mu+ a_{\mathrm{i}}\partial^i A_l  - R_{0 {\mathrm{i}} 0 {\mathrm{l}}}  A^i 
        - (a_{\mathrm{i}}+R_{0 {\mathrm{i}} 0 {\mathrm{m}}} x^{m}) \partial_lA^i\nonumber\\&\:\:\:\:\:\:\:\:\:\:\:\:\:\:\:\:\:\:\:\:\:\:\:\:\:\:\:\:\:\:\:\:\:\:\:\:\:\:\:-\frac{2}{3} R_{\textrm{lj}}A^j + \frac{2}{3} R_{\textrm{l}0}A_\tau + \frac{1}{3} R_{0\textrm{l}0\textrm{j}}A^j+\frac{1}{3}R\indices{_{0 i 0 m}}x^m\partial^i A_l
        -\frac{2}{3}R_{ {\mathrm{i}}  {\mathrm{m}}} x^{m} \partial^i A_l = - 4\pi j_l.\label{Aspace}
    \end{align}
    \end{widetext}

    It is worth pointing out that a similar result is computed in \b \cite{ParkerRevLett,parker} \c using the Lorenz gauge in the absence of acceleration. Overall, Eq. \eqref{Atime} and \eqref{Aspace} can be used to compute the electromagnetic potential in a given reference frame, provided the sources are sufficiently close to the origin \b of the Fermi normal coordinates. This condition translates to the statement that the proper distance between the charge and the curve $z(\tau)$ is much smaller than the acceleration of the curve and the curvature of spacetime along the trajectory. \c This would be the case for a pointlike charge moving along the trajectory used for the expansion, for example.
    
\subsection{An atom in curved spacetimes}\label{subHydrogen}

    As a first explicit example of the formalism developed in Section \ref{secSchInCur}, in this subsection we study the case of an atom under the influence of gravity in a \b general  spacetime that fits the regularity conditions discussed in Section \ref{secSpniCur}\c. We then specialize to the case of Schwarzschild spacetime. A physical situation that could be described by this is an atom on the surface of Earth or in the vicinity of a black hole. \textcolor{black}{This example has been studied in the literature before, see \cite{parker, ParkerRevLett,ParkerHydrogen,ZhaoHydrogen,ZhaoHydrogen2,accHydro}. In this manuscript, we generalize these results by using the Hamiltonian in Eq. \eqref{YES}, which combines the first order in acceleration and curvature contributions. The Hamiltonian in Eq. \eqref{YES} also contains novel first order in curvature terms that have not been analyzed in the references \cite{parker, ParkerRevLett,ParkerHydrogen,ZhaoHydrogen,ZhaoHydrogen2} due to the fact that their contribution is usually $10^4$ times smaller than the other terms.}
    
    To consider an atom, we assume the nucleus to be localized in the center of mass of the system and to have a mass much larger than the electron's. We can then consider the nucleus as a pointlike positive charge that undergoes the trajectory $z(\tau)$. Its coordinates in the FNC associated with this trajectory are then $\bm x = 0$. We can then solve for the first order corrections to the electromagnetic potential $A_\mu$ associated with this source in the Coulomb gauge.  
    The zeroth order term is given by the simple solution of an inertial charge. Namely, we have
    \begin{align}
        A^{(0)}_\tau = - \frac{Ze}{r},&&& A^{(0)}_i = 0,
    \end{align}
    where $e$ is the fundamental charge, $Z$ is the atomic number of the atom and $r = \sqrt{x_ix^i}$. Next, we will solve Eqs. \eqref{Atime} and \eqref{Aspace} for a first order correction to the potential. We write
    \begin{equation}
        A_\mu = A_\mu^{(0)} + A_\mu^{(1)}
    \end{equation}
    and solve for $A_\mu^{(1)}$ only considering first order terms in acceleration and curvature. The solution reads    
    \begin{align}
        A_{0}^{(1)}=& \frac{1}{12} Z e\left(R+4 R_{00}\right) r- Ze\frac{a_\mathrm{i}x^i}{2r}\nonumber\\
        &+\frac{1}{12} Z e\left(3 R\indices{^{0}_{\mathrm{l 0 m}}}-R_{l m}\right) x^{l} x^{m} r^{-1}\label{A01}\\
        A_{i}^{(1)}&=\frac{1}{2} Z e R_{0\mathrm{i}} r+\frac{1}{6} Z e R\indices{^{0}_{\mathrm{l i m}}} x^{l} x^{m} r^{-1}.\label{Ai1}
    \end{align}   
    
    Thus, the full Hamiltonian that contemplates the coupling with electromagnetism for an electron in an atom can be computed from Eqs. \eqref{A01} and \eqref{Ai1}. To first order in acceleration and curvature, we obtain
    \begin{align}\label{HEMatom}
        \mathcal{H}_{EM} =& -\frac{Ze^2}{r}+ Z e^2\left(3 R\indices{^{0}_{\mathrm{l 0 m}}}-R_{\mathrm{l m}}\right) \frac{x^{l} x^{m}}{12r}\nonumber\\
        &+\frac{1}{12} Z e^2\left(R+4 R_{00}\right) r  -Ze^2\frac{a_\mathrm{i}x^i}{2r}\\
        &+i\frac{Ze^2}{2m}\left( R_{0\mathrm{i}}\frac{x^i}{3r}+ r R_{0 \mathrm{i}}\partial^i+ R\indices{^{0}_{\mathrm{l i m}}} \frac{x^{l} x^{m}}{3r}\partial^i\right).\nonumber
    \end{align}
    Together with Eq. \eqref{YES}, it is possible to obtain the energy corrections to first order in acceleration and curvature. Using standard techniques from non relativistic quantum mechanics, we can compute the energy corrections and the change of the atomic orbitals.
    
    We then specialize to the case of a static atom in Schwarzschild spacetime. This could describe numerous phenomena, such as an atom on the surface of a planet or in the vicinity of an irrotational black hole. The metric can be written in coordinates $(t,r,\theta,\phi)$ as
    \begin{equation}\label{gSchw}
        g = -f(r)dt^2 + \frac{dr^2}{f(r)} + r^2(d\theta^2+\sin^2\theta d\phi^2)
    \end{equation}
    with the radial metric function $f(r)$ given by
    \begin{equation}
        f(r) = 1-\frac{2M}{r}.
    \end{equation}
    $M$ can be seen as the mass of the spherically symmetric distribution of matter that generates the gravitational field.
    
    The first step towards applying the formalism of wavefunctions associated with a trajectory in curved spacetimes is to fix the trajectory of the center of mass, $z(\tau)$. We are interested in the description of a static atom in the geometry \eqref{gSchw}. Thus, assuming the atom to be located at a fixed radius $r_0$, the trajectory can be parametrized as
    \begin{equation}
        z(\tau) = \left(\frac{\tau}{\sqrt{f_0}}, r_0,\frac{\pi}{2} ,0\right),
    \end{equation}
    where $\tau$ is the proper time of the trajectory and we define $f_0 = f(r_0)$. Its four velocity is then parallel to the timelike vector $\partial_t$, and is explicitly given by
    \begin{equation}
        u = \frac{1}{\sqrt{f_0}}\partial_t.
    \end{equation}
    
    The next step is to compute the acceleration and curvature along the curve $z(\tau)$ in the Fermi-Walker frame. This frame can be easily defined by normalizing the coordinate vectors. That is, along the worldline we define
    \begin{align}
        e_0 &= u,&&& e_1 &= \sqrt{f_0}\partial_r,\\
        e_2 &= \frac{1}{r_0}\partial_\theta, &&& e_3 &= \frac{1}{r_0} \partial_\phi.\nonumber
    \end{align}
    These can be shown to satisfy Eq. \eqref{FWtransp} along $z(\tau)$. The non-vanishing components of the Riemann tensor in this frame can then be written as
    \begin{align}
        R_{0101} &= -\frac{2M}{r_0^3}, &
        R_{0202} &= \frac{M}{r_0^3}, &  R_{0303} &= \frac{M}{r_0^3}, \\ R_{1212} &= -\frac{M}{r_0^3}, & R_{1313} &= -\frac{M}{r_0^3}, & R_{2323} &= \frac{2M}{r_0^3}. &\nonumber
    \end{align}
    Due to the fact that the metric \eqref{gSchw} describes vacuum solutions to Einstein's equations, both the Ricci tensor and scalar vanish. The acceleration of the curve is given by
    \begin{equation}
        a = \frac{Du}{\dd\tau} = \frac{1}{\sqrt{f_0}}\frac{M}{r_0^2}e_1,
    \end{equation}
    that is, static observers have a constant acceleration in the positive radial direction.
    
    In the reference frame of the atom it is useful to label the FNC according to $x^1 = z$, $x^2 = x$, $x^3 = y$, defining $\bm r = (x,y,z)$. Although at first unhinged, this convention makes the $z$ axis orthogonal to the spherical shells in Schwarzschild spacetime. The electromagnetic coupling Hamiltonian in Eq. \eqref{HEMatom} then yields
    \begin{align}
        \mathcal{H}_{EM} =& -\frac{Ze^2}{r} - \frac{Z e^2}{2r} \frac{M}{r_0^2} \frac{z}{\sqrt{f_0}} \\&-\frac{Z e^2}{r}\frac{ M}{2r_0^3}z^2 + \frac{Z e^2}{r}\frac{ M}{4 r_0^3}(x^2+y^2)\nonumber,
    \end{align}
    where we see the Coulomb term, acceleration correction term and curvature correction terms, respectively. In the expression above $r = \sqrt{x^2+y^2+z^2}$, while $r_0$ is the radial coordinate of the atom in the spacetime. When we put the result above with the Hamiltonian of Eq. \eqref{YES}, we obtain the full Hamiltonian of the system,
    \begin{align}\label{Hhydro}
        \mathcal{H} =& \:\underbrace{m -\frac{Ze^2}{r}-\frac{\nabla^2}{2m}}_{\mathcal{H}_0} +\H_m+\H_e +\H_k,
    \end{align}
    where we define
    \begin{align}   \label{Hm}  
        \H_m=m&\frac{M}{r_0^2}\left(\frac{r^2}{2r_0}-\frac{3z^2}{2r_0}+\frac{z}{\sqrt{f_0}}\right),\\
        \H_e =  \frac{Z e^2}{r} &\frac{M}{r_0^2} \left( \frac{r^2}{4r_0}-\frac{3z^2}{4r_0}-\frac{z}{2\sqrt{f_0}}\right),\label{He}
    \end{align}
    and
    \begin{widetext}
    \begin{align}
        \H_k = &-\left(\frac{3}{4m}\frac{M}{r_0^2}\frac{z}{\sqrt{f_0}} + \frac{3}{8m}\frac{M}{r_0^3}r^2-\frac{9}{8m}\frac{M}{r_0^3}z^2\right)\nabla^2  -\left(\frac{3}{4m}\frac{M}{r_0^2}\frac{\partial_z}{\sqrt{f_0}} + \frac{5}{12m}\frac{M}{r_0^3}\bm r \cdot \nabla-\frac{5}{4m}\frac{M}{r_0^3}z\partial_z\right)\\
        &\:\:\:\:\:+ \frac{1}{6m}\frac{M}{r_0^3}\left((x^2 + y^2)\partial_z^2 + z^2 (\partial_x^2+\partial_y^2) - 2 x^2 \partial_y^2 - 2 y^2 \partial_x^2 - 2 xz\partial_x \partial_z - 2 yz \partial_y \partial_z + 4 xy \partial_y\nonumber \partial_x\right)
    \end{align}
    \end{widetext}
    %\begin{align}
        %\mathcal{H} =& \:m -\frac{\nabla^2}{2m} -\frac{Ze^2}{r}+\frac{m}{2}\frac{M}{r_0^3}r^2-\frac{3m}{2}\frac{M}{r_0^3} z^2+m \frac{M}{r_0^2}\frac{z}{\sqrt{f_0}}- \frac{Z e^2}{r} \frac{M}{r_0^2} \frac{z}{\sqrt{f_0}} -\frac{Z e^2}{r}\frac{ M}{2r_0^3}z^2 + \frac{Z e^2}{r}\frac{ M}{4 r_0^3}(x^2+y^2)\\
        %&-\left(\frac{3}{4m}\frac{M}{r_0^2}\frac{z}{\sqrt{f_0}} + \frac{3}{8m}\frac{M}{r_0^3}r^2-\frac{9}{8m}\frac{M}{r_0^3}z^2\right)\nabla^2  -\left(\frac{3}{4m}\frac{M}{r_0^2}\frac{\partial_z}{\sqrt{f_0}} + \frac{3}{4m}\frac{M}{r_0^3}\bm r \cdot \nabla-\frac{9}{4m}\frac{M}{r_0^3}z\partial_z\right)\\
        %&-\frac{1}{3m}\frac{M}{mr_0^3}(z^2(\partial_x^2+\partial_y^2)-(x^2+y^2)\partial_z^2)+\frac{M}{r_0^3}(x^2\partial_y^2+y^2 \partial_x^2)+\frac{1}{3m}\frac{M}{r_0^3}\bm r \cdot \nabla -\frac{M}{m r_0^3} z\partial_z.\nonumber
    %\end{align}
    Notice that $\mathcal{H}_0$ is the hydrogen atom Hamiltonian, with the rest mass term. The remaining terms $\mathcal{H}_m$, $\mathcal{H}_e$ and $\mathcal{H}_{k}$ are the correction terms due to the curvature of spacetime and acceleration. Notice that $\mathcal{H}_m$ is proportional to the electron mass, $\mathcal{H}_e$ is proportional to the squared electron charge, $e^2$, and $\mathcal{H}_{k}$ is associated with the corrections to the kinetic term. Under the assumption that these three terms are small compared to $\mathcal{H}_0$, an expansion can be performed to compute the shift in the energy levels and eigenfunctions.
    
    We proceed to analyze the typical orders of magnitude of each of the terms in Eq. \eqref{Hhydro}. We will work with natural units, so that the Bohr radius is of the order of $a_0 \sim 10^{24}\ell_p$, where $\ell_p$ is the Planck length. Then, the following relation holds between the mass of the electron $m$, the electron charge $e$ and the fine structure constant $\alpha$,
    \begin{equation}
        m a_0 = \frac{1}{\alpha} = \frac{1}{e^2}.
    \end{equation}
    The relation above is important because all the approximations involved will regard the relationship of $m$ and $a_0$ with the Schwarzschild radius $r_S = 2M$ and the coordinate radius $r_0$. As a matter of fact, we can quantify the order of magnitude of each of the terms in the Hamiltonian above. For that, we use the fact that the expected values of position and momentum are of the order of $\ev{r} \sim a_0$ and $\ev{\partial}\sim a_0^{-1}$. We can then estimate the expected values of the following terms
    \begin{align}\label{ordersAtom}
        m \sim \frac{1}{\alpha a_0},&&&-\frac{Ze^2}{r} \sim \frac{\alpha}{ a_0},&&&-\frac{\nabla^2}{2m} \sim  \frac{\alpha}{a_0}.
    \end{align}
    With these we can compute the order of magnitude of the hydrogen Hamiltonian, 
    \begin{equation}\label{orderH0}
        \mathcal{H}_0 \sim \frac{1}{a_0}\left(\frac{1}{\alpha} + \alpha\right),
    \end{equation}
    where the first term is associated with the rest mass and the second to the hydrogen atom energy levels. Thus, the rest mass of the electron is of the order of $\alpha^{-2}\sim 10^{4}$ times larger than the energy levels of the electron in the atom. 

    %To compute the order of magnitude of the terms $\H_m, \H_e$ and $\H_k$, 
    The acceleration and curvature for the static trajectory chosen are of order
    \begin{align}\label{ordersSchw}
        a \sim \frac{r_S}{r_0^2}, &&& R \sim \frac{r_S}{r_0^3}.
    \end{align}
    We thus conclude that the orders of magnitude of the correction terms are
    \begin{align}
        \H_m \sim&\label{HmEst} \frac{1}{\alpha} \frac{r_S}{r_0^2}\left(1+\frac{a_0}{r_0}\right),\\
       \H_k\sim \H_e \sim&\label{HeEst} \:{\alpha} \frac{r_S}{r_0^2}\left(1+\frac{a_0}{r_0}\right),
    \end{align}
    where the $a_0/r_0$ term in the parenthesis is the contribution due to curvature, while the other is due to acceleration. We can then see that the curvature correction tends to be orders of magnitude less than the acceleration terms \b whenever $a_0\ll r_0$. Noticing that in order for the atom to be outside of the black hole, one must have $r_0>r_S$, we obtain that the curvature will contribute less than acceleration whenever $r_S > a_0$, which is a reasonable assumption for the black hole radius. \c Another conclusion is that the $\H_m$ term is of the order of $\alpha^{-2}\sim 10^{4}$ larger than the $\H_e$ and $\H_k$ corrections. The full correction Hamiltonian $\H_I = \H_m + \H_e + \H_k$ is then of order
    \begin{align}\label{orderHI}
        \mathcal{H}_I \sim \left(\frac{1}{\alpha} +\alpha\right)\frac{r_S}{r_0^2}\left(1+\frac{a_0}{r_0}\right),
        %\mathcal{H} %=& m +\frac{1}{ma_0^2}+\frac{1}{a_0}+\frac{mMa_0}{r_0^2}\left(\frac{a_0}{r_0}+1\right)+\frac{M}{r_0^2}\left(\frac{a_0}{r_0}+1\right)\\
        %&+\frac{Ma_0}{mr_0^2}\left(\frac{a_0}{r_0}+1\right)\frac{1}{a_0^2}+\frac{M}{m r_0^2}\frac{1}{a_0}\left(\frac{a_0}{r_0}+1\right)\\
        %=& m +\frac{\alpha}{a_0}+\frac{1}{a_0}+\frac{M}{\alpha r_0^2}\left(\frac{a_0}{r_0}+1\right)+\frac{M}{r_0^2}\left(\frac{a_0}{r_0}+1\right)\\
        %&+\frac{M\alpha}{r_0^2}\left(\frac{a_0}{r_0}+1\right)
    \end{align}
    where the respective contributions of $\H_m, \H_e$ and $\H_k$ have been factored.
    
    In order to apply perturbative techniques, we must make sure that $\H_I$ is much smaller than $\H_0$. Comparing Eqs. \eqref{orderHI} and \eqref{orderH0} we see that the condition that is compatible with this approximations is
    \begin{equation}
        \frac{a_0 r_S}{r_0^2}\ll 1.
    \end{equation}
    And indeed, this condition is satisfied unless the Schwarzschild radius of the spacetime is of the order of the Bohr radius and the atom is close to the horizon. Thus, the formalism can be used in most physical situations of atoms on the surface of planets and stars.
    
   %\textcolor{black}{The curvature corrections to energy can then be computed using the techniques of perturbation theory in quantum mechanics. It is important to take under consideration the fact that there are curvature corrections to the inner product, given by Eq. \eqref{innerProdSigmaExp}. We proceed our analysis using the techniques presented in \b \cite{ParkerRevLett,parker}\c for degenerate perturbation theory when perturbations to the  inner product also have to be considered.} 
   We thus compute the corrections due to the inner product, $\H_m$, $\H_e$ and $\H_k$ to the $1s,2s$ and $2p$ energies. We found that the first order coupling does not produce any energy correction to the $1s$ state. The $2s$ and $2p$ states are degenerate, so that we apply the degenerate perturbation theory developed in \b \cite{ParkerRevLett,parker} \c that takes into account the curvature corrections to the inner product. The acceleration and curvature break part of the degeneracy, yielding the following energy corrections for the $n=2$ states
    %\begin{align}
        %E_{2,1,\pm 1} =& \frac{Ma_0}{r_0^3}\left(\frac{4}{\alpha}+\frac{5Z}{6}-\frac{\alpha}{56}\right),\\
        %E^{\pm}_{2} =& \pm\frac{M}{r_0^2 \sqrt{f_0}}\left(\frac{3}{\alpha}+\left(\frac{3}{16}-\frac{Z}{4}\right)\alpha\right)\\&+ \frac{M a_0}{r_0^3}\left(\frac{4}{\alpha}-\left(\frac{219}{280}+\frac{5Z}{6}\right)\alpha\right).
        %\label{124}
    %\end{align}
    %\textcolor{red}{New Result}
    \textcolor{black}{
    \begin{align}
        &E_{2,1,\pm 1} = \frac{r_S a_0}{2r_0^3}\left(\frac{6}{\alpha}+\left(\frac{Z}{2}+\frac{13}{24}\right)\alpha\right),\\
        &\:\:\:\:E^{\pm}_{2} = -\frac{r_S  a_0}{2 r_0^3}\left(\frac{6}{\alpha}+\left(\frac{Z}{2}+\frac{13}{24}\right)\alpha\right)\pm \frac{r_s}{96 r_0^2 \alpha \rho_0} \sqrt{\Delta},\nonumber
        \label{124}
    \end{align}
    where $\Delta$ is a dimensionless parameter given by
    \begin{align}
        \Delta = 9 (48 \!+ \!(3\! -\! 4 Z) \alpha^2)^2 \!+ \!
  \frac{4a_0^2\rho_0^2}{r_0^2} (144 \!+\! (13 + 12 Z) \alpha^2)^2.
    \end{align}}%
    The energy corrections to the states $\psi_{2,1,\pm 1}$ still do not break their degeneracy, and are given by $E_{2,1,\pm 1}$. The degeneracy in the space generated by $\psi_{2,0,0}$ and $\psi_{2,1,0}$ is broken and the energy corrections to the corresponding eigenstates are $E_2^{\pm}$. 
    %We can then see that the mass and electromagnetic terms do not contribute to the energy of the spherically symmetric states. This can be seen from \eqref{Hm} and \eqref{He}. Notice that the correction terms shift the energy levels of the $\psi_{1,0,0}$, $\psi_{2,0,0}$ and $\psi_{2,1,0}$ states down, while it increases the energy of the non $z$ axis-symmetric states $\psi_{2,1,\pm 1}$. Additionally, all contributions to the shifts in the energy levels are due to curvature and not acceleration. Indeed, the expressions in \eqref{124} agrees with the estimates argued in \eqref{HmEst}, \eqref{HeEst} and \eqref{HkEst}, except for the fact that they lack the acceleration contributions.
    \textcolor{black}{This result is compatible with previous calculations in \cite{accHydro}, where only acceleration was considered. It is also compatible with \cite{parker,ParkerRevLett,ParkerHydrogen,ZhaoHydrogen,ZhaoHydrogen2}, where only the $\mathcal{H}_m$ curvature contributions were taken into account in a non-relativistic description. Our result contains additional first order in curvature corrections which, to the authors knowledge, have not been computed before.}
    
\begin{figure}
        \includegraphics[scale=0.8]{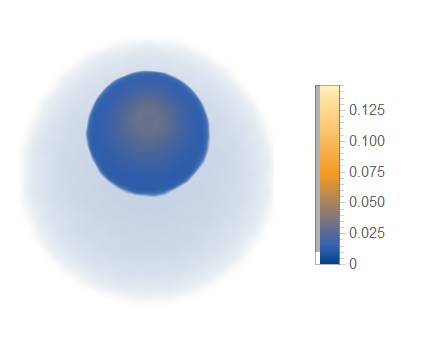}
        \caption{The probability distribution for a hydrogen atom according to Eq. \eqref{plot}. One clearly sees the orbital being dragged down by an upwards pointing acceleration. We have used natural units, and chosen $r_S = 6\times 10^{-8} a_0$, $r_0 = 50r_S$.}
        \label{fig}
    \end{figure}

    It is arguably more interesting to look at the corrections to the wavefunctions themselves, where acceleration plays a crucial role. We will do so for the $1s$ state in order to see the break of the spherical symmetry of the orbitals. Given that the acceleration terms in general contribute more than curvature, we compute the first order corrections considering only a uniformly accelerated atom and neglecting the spacetime curvature. Notice that the theory of perturbation for states in quantum mechanics requires one to compute expected values for all eigenstates of the Hamiltonian. However, it is possible to show that only the $\psi_{n,2k+1,0}$ states contribute non-trivially, for natural $n$ and integer $k$. The highest contribution will then be associated with the $\psi_{2,1,0}$ state. Taking only this contribution into account, one finds
    \begin{align}
        \psi_{1,0,0}(x) = \frac{e^{-r/a_0}}{\sqrt{\pi}a_0^{3/2}} - \frac{128}{729\, \alpha^2}\frac{r_S}{r_0^2 \sqrt{f_0}} \frac{z e^{-r/2a_0}}{\sqrt{\pi}a_0^{3/2}}.\label{plot}
    \end{align}
    We then see a break of the spherical symmetry of the ground state that shifts the probability density down, as can be seen in Fig. \ref{fig}. It is an effect that drags the orbitals in the opposite direction to the acceleration required to stay at rest in this gravitational field. This effect is valid in general spacetimes due to the fact that in most scenarios curvature will contribute orders of magnitude less than acceleration. In fact, Eq. \eqref{plot} neglects curvature effects and thus does not depend on the geometry of spacetime.
    
\subsection{A fermion in AdS spacetime}
    
    In the previous sections we have considered a localized fermionic system in curved spacetimes. In particular, we have studied in detail what happens with an electron bound to an atom, where the Coulomb potential associated with the proton was responsible for localizing the system. However, as we will show in this subsection, one does not require an electromagnetic field to produce localized states. Indeed, looking at Eq. \eqref{YES} and the previous analysis, we see that the highest order contribution in curvature and acceleration is, in principle, given by
    \begin{equation}\label{reads}
        \mathcal{H} = \:m-\frac{\nabla^2}{2m} +m a_{\textrm{j}}x^j+\frac{m}{2}R_{\textrm{k}0\textrm{m}0}x^kx^m.
    \end{equation}
    However, the Laplace operator above is not symmetric with respect to the curved inner product given in Eq. \eqref{innerProdSigmaExp}. In order to make the Hamiltonian Hermitian, one must replace $\nabla^2$ by \textcolor{black}{$\delta^{ij}\hat p_i \hat p_j$, with $\hat p_i$} given by Eq. \eqref{p}, or, equivalently, symmetrize it. \textcolor{black}{Up to a constant factor,} the Hamiltonian can then be written in operator form as
    \color{black}
    \begin{equation}
        \hat{\mathcal{H}} = m + \frac{1}{2m}\delta^{ij}\hat{p}_i\hat{p}_j +m a_{\textrm{j}}\hat{x}^j+\frac{m}{2}R_{\textrm{k}0\textrm{m}0}\hat{x}^k\hat{x}^m.
    \end{equation}
    \color{black}
    %\begin{align*}
        %&\int \psi^* (1-A_{ij}x^i x^j) \partial_k\partial^k\psi \\
        %&= -\int \partial_k\psi^* (1-A_{ij})x^i x^j \partial^k\psi+2\int \psi^* A_{kj} x^j \partial^k\psi\\
        %&=\int \partial^k\partial_k\psi^* (1-A_{ij}x^i x^j) \psi-2\int \partial^k\psi^* (1-A_{ij}x^i x^j)A_{kj} x^j \psi\\
        %&-2\int \psi^* (1-A_{ij}x^i x^j)A_{kj} x^j \partial^k\psi
    %\end{align*}
    
    %\begin{equation}
        %-\nabla^2 \mapsto -\nabla^2 + 2A_{kj}x^j \partial^k
    %\end{equation}
    %\begin{equation}
        %\hat{p}_i = -i \partial_i + i A_{ij}x^j
    %\end{equation}
    %\begin{equation}
        %p_i p^i = - \nabla^2 +A_{i}{}^i +2A_{ij}x^j \partial^i 
    %\end{equation}

    The quadratic dependence on the curvature term in Eq. \eqref{reads} implies that it might be possible to obtain an effective \b harmonic oscillator \c binding potential. Then, in principle, curvature could be responsible for the creation of bound states. In this section we investigate whether our formalism allows for gravitational fields to localize fermionic systems.
    
    In order for curvature to localize a quantum system, it is natural to expect the spacetime to be negatively curved. \b This intuition can be obtained, for instance, from the geodesic deviation equation where negative curvature is responsible for decreasing the distance between neighbouring geodesics~\cite{Wald1,poisson}. An explicit example where negative curvature produces bound states is the case of the standard quantization of the Klein-Gordon field in anti-de Sitter (AdS) spacetime, where the solutions are localized and the corresponding energy spectrum is discrete~\cite{harlow}. In fact, AdS is the simplest spacetime that can be used in order for curvature to produce bound states, for it is a constant curvature manifold~\cite{hawking}\c. In particular, the Riemann curvature tensor is given by
    %In the previous sections, we considered the coupling of an atom to the electromagnetic field. The attractive Coulomb interaction lead to bound states. We could compute the energy corrections due to curvature and acceleration.
    
    %In the following we study an uncharged particle in a pure gravitational field and look in detail at the interaction of the particle with gravity. It turns out that the background curvature of spacetime can lead to the formation of bound states. Such states exist in the example of anti-de Sitter spacetime. 
    
    %The anti-de Sitter spacetime is a constant curvature spacetime with the Riemann curvature tensor given by 

    \begin{align}\label{cteR}
        R_{\mu \nu \rho \sigma} = -\frac{1}{\ell^2}\qty(g_{\mu \rho} g_{\nu \sigma} - g_{\mu \sigma} g_{\nu \rho}),
    \end{align}
    where $\ell$ is the curvature radius.
    
    In this spacetime, we shall consider a quantum mechanical system undergoing a trajectory $z(\tau)$. We assume to have a frame which is Fermi-Walker transported along this trajectory and associate Fermi normal coordinates to the worldline. Notice that in constant curvature spacetimes, such as the anti-de Sitter, the components of the curvature tensor in any orthonormal frame are the same. This can be seen from Eq. \eqref{cteR}. We then have
    \begin{align}
        R_{\mathrm{k} 0 \mathrm{m} 0} = -\frac{1}{\ell^2} \eta_{\mathrm{k}\mathrm{m}} \eta_{00} = \frac{1}{\ell^2} \delta_{\mathrm{k}\mathrm{m}}.
    \end{align}   
    Plugging the result above in Eq. \eqref{reads}, we obtain the Hamiltonian
    \b\begin{align}\label{binding}
        \hat{\mathcal{H}} =& \:m+m\,\bm{a}\cdot \hat{\bm x} +\frac{m}{2\ell^2}\hat{\bm x}^2+\frac{\hat{\bm p}^2}{2m},
    \end{align}\c
    \textcolor{black}{where we use bold symbols to denote spacelike vectors tangent to the system's rest space and $\:\cdot\:$ denotes the flat inner product.}
    
    Thus, in Eq. \eqref{binding} the curvature of spacetime acts as an effective potential for the particle. The contribution is quadratic in the FNC $\bm x$, thus we expect some bound states similar to a harmonic oscillator. Indeed, we can rewrite the Hamiltonian as \b
    \begin{align}
        \hat{\mathcal{H}} =& \:m +\frac{m}{2\ell^2} \qty(\hat{\bm{x}} + \ell^2 \bm{a})^2 - \frac{m \ell^2}{2}\bm a ^2+\frac{\hat{\bm p}^2}{2m}.
    \end{align}\c
    It is easy to see that under the transformation \mbox{$\hat{\bm y} = \hat{\bm x} + \ell^2 \bm a$}, the Hamiltonian transforms into a three-dimensional Harmonic oscillator with a constant shift in energy. 
    %\begin{align}
        %\mathcal{H} =& \:m +\frac{m}{\ell^2} \bm y^2 -\frac{\nabla_y^2}{2m}- \frac{m}{4\ell^2}\bm a^2.
    %\end{align}
    We can then read off the frequency of the effective harmonic oscillator 
    \b\begin{equation}
        \omega^2 = \frac{1}{\ell^2}.
    \end{equation}\c
    %This immediately gives the corresponding energy eigenvalues of the Hamiltonian
    %\begin{align}
        %E_{n_1,n_2,n_3} = \omega \qty(n_1 + n_2 + n_3 + \frac{3}{2}) + m - \frac{m}{4\ell^2} a_i a^i.
    %\end{align}
    
    The corresponding eigenfunctions are shifted, such that their centre is at $\bm x_0 = -\ell^2 \bm a$. This means that due to the acceleration correction, the wavefunction is moved in the direction opposite to $\bm a$, as if the wavefunction was lagging behind, similar to what we saw in the previous Subsection \ref{subHydrogen}.
    
    In order for the potential of Eq. \eqref{reads} to trap the fermionic particle we are describing, it must be of the same order of magnitude as the kinetic energy of the system. To study whether this is compatible with our expansions, we first assume the system to start in a state that is localized around a region of characteristic size $\sigma$. 
    With this we can estimate the order of magnitude of the terms in Eq. \eqref{reads} by using
    \begin{align}\label{sim}
        \bm x^2 \sim \sigma^2 &&& \bm p^2 \sim \sigma^{-2}.
    \end{align}
    In order to have bound states, the potential term must be of the same order of magnitude as the kinetic term. Imposing this condition yields the following estimate for the localization of the system in terms of its mass and the spacetime curvature
    \begin{equation}\label{ARGH}
        \frac{m}{\ell^2}\sigma^2 \sim \frac{1}{m \sigma^2}\Rightarrow \sigma \sim \left(\frac{\ell}{m}\right)^{1/2}.
    \end{equation}
    The equation above then tells us that, in principle, it is possible to localize the system in a region of size of the order of $(\ell/m)^{1/2}$. Notice that this quantity is the geometric mean between the Compton wavelength of the particle and the curvature radius. 
    
    This result is also compatible with the previous expansion we have used, where we assumed curvature to be small. Indeed, the localization of the particle, $\sigma$, has to be larger than the Compton wavelength $m^{-1}$. In Eq. \eqref{ARGH}, we then obtain the condition $(m\ell )^{-1}\leq 1$, while in our expansions we assume to have
    $R_{0 \textrm{i} 0 \textrm{j}}x^i x^j\ll 1$. This translates into $(m \ell)^{-1}\ll 1$ when one uses Eqs. \eqref{sim} and \eqref{ARGH}. It is important to remark that in this example the spacetime curvature would need to be very large compared to most astrophysical phenomena, but our expansions would still be valid due to the small size of the system. 
    
    %We now justify the claim made earlier in the subsection, where we stated that the correction that would make the Hamiltonian self-adjoint could be neglected for the purpose of this example. In order for curvature to create bound states, we studied the case where the kinetic term is of the same order of magnitude as the potential produced by curvature. Thus, any correction to the kinetic term would be of second order in curvature and can be neglected in our expansions.
    
    In summary, we conclude that curvature itself might be a mechanism to produce localized states of fermionic particles. The localization produced by weak gravitational fields would then be able to localize particles in regions larger than their Compton wavelength, but much smaller than spacetime curvature.

\section{Conclusion}\label{secConclusion}

    We have provided the means by which one can use a Schr\"odinger complex wavefunction description for a fermionic \tb{particle} undergoing an arbitrary trajectory in curved spacetimes. Our results are valid provided that the acceleration and curvature along the center of mass worldline are small compared to the size and rest mass of the system. We were able to provide \tb{a Hilbert space,} an inner product and obtain a Hamiltonian operator (given in Eq. \eqref{YES}) associated with the proper energy of the system. The Hamiltonian then generates a unitary time evolution for complex wavefunctions defined in the center of mass rest spaces.
    
    Explicitly writing down the FNC around a given spacetime in terms of other coordinate systems tends to be an unpractical task. However, our description does not require one to relate the FNC with any other coordinate system. Instead, the formulation happens with respect to the rest frame of the observer. This means that, for all purposes, the description can be done as if one was applying the techniques of non-relativistic quantum mechanics in Cartesian coordinates with external force terms. These terms could then be calculated in terms of the acceleration and curvature along the system's trajectory.
    
    To apply this formalism it is then only required to have the Riemann curvature tensor and acceleration along the center of mass worldline. With this, writing these tensors in the Fermi-Walker frame associated with the curve yields all the necessary tools for calculating expected values of observables. The theory yields the correct predictions for expected values to first order in curvature and acceleration, provided one neglects the spin degree of freedom of the system. This is compatible with the single component complex wavefunction description presented here. If instead one wants to consider fully relativistic quantum mechanics and Dirac's equations, we also provide a general form for the Hamiltonian expanded to first order in acceleration and curvature \tb{in Eq. \eqref{HDiracExp}}.
    
    Moreover, we present corrections for the coupling of a \tb{one-particle} quantum system with electromagnetism in the presence of acceleration and curvature. The acceleration terms could be relevant if, for example, one wishes to probe the Unruh effect by accelerating a probe \b \cite{Unruh1976,Unruh-Wald,unruhEffectNoThermal,unruhSlow,vanzella}\c. In these extreme regimes the correction terms we found might produce relevant effects to its observation. In particular we have seen that the description of a static atom in a spherically symmetric gravitational field depends on both acceleration and curvature. Indeed, both contribute to the correction terms that arise in this formulation.
    
    Our results also provide means by which one can study the coupling of a nonrelativistic quantum \tb{particle} with gravity. Possible applications of these studies range from using atoms to probe spacetime curvature \cite{parker,topology,prlInterferometer,exp1,exp2} to probing the quantum nature of the gravitational field \cite{remi}. \color{black}
    Overall we conclude that the Hamiltonian in Eq. \eqref{YES} with the inner product \eqref{innerProdSigmaExp} provide the description of a Schr\"odinger wavefunction in curved spacetimes. This allows one to compute the leading order corrections to the eigenstates and eigenvalues of any \tb{one-particle} wavefunction system due to spacetime curvature and acceleration. In most regimes it can be reduced to the following Hamiltonian
    \begin{equation}
        \mathcal{H} = \mathcal{H}_0 + m a_{\textrm{j}}x^j+\frac{m}{2}R_{\textrm{k}0\textrm{m}0}x^kx^m,
    \end{equation}
    where $\mathcal{H}_0$ is the original Hamiltonian in its reference frame, $a$ is the proper acceleration of the system and $R$ the Riemann curvature tensor evaluated along the worldline. 
    \color{black}
    
    Notice that although in this work we have considered both the electromagnetic and gravitational fields to be classical, the description could be extended to couple systems to quantum fields. Our formalism would then describe a localized nonrelativistic quantum system interacting with a quantum field. Systems that satisfy these properties are commonly called particle detector models \b \cite{us,measQFT,delocAtom,us2} \c and they are usually motivated  by considering a Schr\"odinger atom interacting with electromagnetism. Our formalism then allows for this analogy to be studied when the atom undergoes an arbitrary trajectory in curved spacetimes.
    
    Complementary, one can think of a didactic application of this manuscript. The formalism developed here allows for a student who has learned nonrelativistic quantum mechanics to perform calculations regarding quantum \tb{particles} in curved spacetimes. Overall, we have simplified the complex description of a spinor in curved spacetimes to a formalism that could be presented to physics undergraduate students.

\section{Acknowledgements}

    We would like to thank the PSI program for facilitating this research, with special thanks to Gang Xu and Dan Wohns for introducing us to spinors and instigating our pursue of a deeper understanding of spinors in curved spacetimes. We would like to thank Erickson Tjoa and Lars Dehlwes for reviewing the manuscript. The authors also thank Bruno de S. L. Torres, Matheus H. Zambianco and Lars Dehlwes for insightful discussions. \fix{The authors also thank Ashkan Alibabaei for idenitfying minor typos in previous versions of the manuscript, and highlighting that a more detailed explanation would be fit in Appendix \ref{appSchCur}.} \textcolor{black}{Most of this work has been conducted at Perimeter Institute.} Research at Perimeter Institute is supported in part by the Government of Canada through the Department of Innovation, Science and Economic Development Canada and by the Province of Ontario through the Ministry of Economic Development, Job Creation and Trade. T.R.P. thanks Drs. David Kubiznak and  Eduardo Martin-Martinez’s funding through their NSERC Discovery grants. J.N. also thanks the Max Weber Stiftung for financial support.%\textcolor{red}{The authors would like to stress that most of this work was done at Perimeter Institute and does not necessarily reflect the opinions or views of their current affiliations.} 

\appendix

\section{The Frame Connection}\label{appFrame}
    
    The goal of this appendix is to present the computations that lead to Eqs. \eqref{omega1}, \eqref{omega2}, \eqref{omega3}, \eqref{omega4} and \eqref{omega5} for the frame connection coefficients. In order to do that we make use of the following equation
    \b\begin{equation}\label{gammaFrameGammaCoord}
        \omega\indices{_{\mu}^I_J}  = \Gamma_{\mu \nu}^\alpha e^{I}_\alpha e^\nu_J- e^\nu_J \partial_\mu e^I_\nu,
    \end{equation}\c
    for the change of Christoffel symbols from one frame to another. The first step is then to compute the derivatives of the frame. These can be computed by direct calculation and yield
    \begin{align}
        \partial_\mu  e_{\tau}^{I}&=\delta_{0}^{I}a_\mu-\frac{1}{2} (R\indices{^{I}_{\mathrm{k 0 m}}}+\b R\indices{^{I}_{\mathrm{m 0 k}}}\c )\delta^k_\mu x^{m}  \\
        \partial_\mu e_{i}^{I}&=-\frac{1}{6} (R\indices{^{I}_{\mathrm{k i m}}} +R\indices{^{I}_{\mathrm{m i k}}} )\delta^k_\mu x^{m}.
    \end{align}
    With this and the expressions for the Christoffel symbols from Eqs. \eqref{gamma1}, \eqref{gamma2}, \eqref{gamma3}, \eqref{gamma4}, \eqref{gamma5} and \eqref{gamma6}, it is possible to obtain the connection coefficients in the orthonormal frame. These read

      \begin{align}
        \omega\indices{_{\tau}^0_0} =&  0\label{A4}\\
        \omega\indices{_{i}^0_0} =& a_\mathrm{i} +R_{\mathrm{0 i 0 m}} x^m - (a_\mathrm{i}-R\indices{^0_{\mathrm{i0m}}}x^m) = 0\\
        \omega\indices{_\tau^{{\mathrm{i}}}_0} =& a^\mathrm{i}+R\indices{_{0}^{\mathrm{i}}_{0 \mathrm{m}}} x^{m}\\
        \omega\indices{_i^{{\mathrm{j}}}_0}=& R\mathrm{\indices{_{0 m i}^j}} x^m -\frac{1}{2}(R\mathrm{\indices{_{i}^{j}_{0m}}}+R\mathrm{\indices{_{i0}^j_m}})x^m\nonumber \\
        =& \frac{1}{2}(R\mathrm{\indices{_i^j_{0m}}}-R\mathrm{\indices{_{i0}^j_m}})x^m\nonumber\\
        =& \frac{1}{2}(R\mathrm{\indices{_{0mi}^j}}+R\mathrm{\indices{_{0i}^j_m}})x^m\nonumber\\
        =& \frac{1}{2}R\mathrm{\indices{_{0}^j_{im}}}x^m\\
        \omega\indices{_\tau^{0}_{{\mathrm{i}}}} =& a_\mathrm{i} +R\mathrm{_{0 i 0 m}}x^m\\
         \omega\indices{_j^{0}_{{\mathrm{i}}}}=& \frac{1}{3}\left(R\mathrm{_{0 i j m}}+R\mathrm{_{0 j i m}}\right) x^{m}+\frac{1}{6} (R\mathrm{\indices{^{0}_{j i m}}} +R\mathrm{\indices{^{0}_{m i j}}} ) x^{m}\nonumber\\
         =& \frac{1}{3}\left(R\mathrm{_{0 i j m}}+\mathrm{R_{0 j i m}}\right) x^{m}\nonumber\\
         =&-\frac{1}{6} (R\mathrm{\indices{_{0j i m}}} -R\mathrm{_{0ijm}}+R\mathrm{_{0jim}} ) x^{m}\nonumber\\
         =& \frac{1}{2}R\mathrm{_{0 i j m}} x^{m}\\
         \omega\indices{_{\tau}^{{\mathrm{i}}}_{{\mathrm{j}}}} =& R\mathrm{\indices{_{0 m j}^i}}x^m\\
         \omega\indices{_k^{{\mathrm{j}}}_{{\mathrm{i}}}} =& \frac{1}{3}\left(R\mathrm{\indices{_{k}^j_{ i m}}}+R\mathrm{\indices{_{i}^j_{ k m}}}\right) x^{m}+\frac{1}{6} (R\mathrm{\indices{^{j}_{ k i m}}} +R\mathrm{\indices{^{j}_{m i k}}} ) x^{m}\nonumber\\
         =& \frac{1}{3}\left(R\mathrm{\indices{_{i m k}^j}}+R\mathrm{\indices{_{i}^j_{k m}}}\right) x^{m}+\frac{1}{6} (R\mathrm{\indices{_{i m}^j_k}} +R\mathrm{\indices{_{i k}^j_m}} ) x^{m}
        \nonumber \\
         =& \left(\frac{1}{6} R\mathrm{\indices{_{i m k}^j}}+ \frac{1}{3}R\mathrm{\indices{_{i}^j_{k m}}}+\frac{1}{6}R\mathrm{\indices{_{i k}^j_m}}\right) x^{m}\nonumber\\
         =& \frac{1}{2}R\mathrm{\indices{_i^j_{k m}}x^{m}},\label{A11}
    \end{align}
    where we used the Bianchi identity to rewrite the curvature terms in the equations above,
    \begin{equation}
        \begin{gathered}
        R\mathrm{_{0mij}} =-R\mathrm{_{0ijm}}+R\mathrm{_{0jim}},\\
        R\mathrm{\indices{_{i k j m }}}+R\mathrm{\indices{_{i m k j}}} = R\mathrm{_{ijkm}}.
        \end{gathered}
    \end{equation}
    
\section{Dirac's Hamiltonian around a Worldline}\label{appDirH}
    In this appendix we explicitly show the calculations that lead from the exact expression of the Hamiltonian from Eq. \eqref{HDiracFNC} to the expansion in terms of curvature and acceleration from Eq. \eqref{HDiracExp}. The first step is then the computation of the spin connection, given by 
    \begin{align}
        \Gamma_\mu = -\frac{1}{2}\omega_{\mu I J}S^{I J} = -\frac{1}{4}\omega_{\mu I J}\gamma^{I} \gamma^{J}
    \end{align}
    For that purpose, we rewrite the generators of the $SL(2,\mathbb{C})$ action over the spinor bundle in the following form
    \begin{align}
        S^{IJ} = \frac{1}{4}[\gamma^I,\gamma^J] = \frac{1}{2}\gamma^{[I} \gamma^{J]}.
    \end{align}
    Using the expression for the frame Christoffel symbols \b in Eqs. (\ref{A4}--\ref{A11}), we find the following components for the spin connection:
    \begin{align}
        \Gamma_\tau &= -\frac{1}{4}\left( \omega\indices{_{\tau{\mathrm{i}}0}}\gamma^{{\mathrm{i}}} \gamma^{0}+\omega\indices{_{\tau0{\mathrm{i}}}}\gamma^{0} \gamma^{{\mathrm{i}}}+\omega\indices{_{\tau{\mathrm{i}}{\mathrm{j}}}}\gamma^{{\mathrm{i}}} \gamma^{{\mathrm{j}}}\right)\\
         &= -\frac{1}{4}\left( 2\omega\indices{_{\tau{\mathrm{i}}0}}\gamma^{{\mathrm{i}}} \gamma^{0}+\omega\indices{_{\tau{\mathrm{i}}{\mathrm{j}}}}\gamma^{{\mathrm{i}}} \gamma^{{\mathrm{j}}}\right)\nonumber\\
        &=-\frac{1}{2}(a_\mathrm{i}+R\mathrm{_{0i0m}}x^m) \gamma^{{\mathrm{i}}}\gamma^0 \fix{+} \frac{1}{4}R\mathrm{_{0 m ij}}x^m\gamma^{{\mathrm{i}}}\gamma^{{\mathrm{j}}},\\
        \Gamma_{k} &=-\frac{1}{4}\left( \omega\indices{_{k{\mathrm{i}}0}}\gamma^{{\mathrm{i}}} \gamma^{0}+\omega\indices{_{k 0{\mathrm{i}}}}\gamma^{0} \gamma^{{\mathrm{i}}}+\omega\indices{_{k{\mathrm{i}}{\mathrm{j}}}}\gamma^{{\mathrm{i}}} \gamma^{{\mathrm{j}}}\right)\nonumber\\
        &= -\frac{1}{4}\left( 2\omega\indices{_{k{\mathrm{i}}0}}\gamma^{{\mathrm{i}}} \gamma^{0}+\omega\indices{_{k{\mathrm{i}}{\mathrm{j}}}}\gamma^{{\mathrm{i}}} \gamma^{{\mathrm{j}}}\right)\nonumber\\
        &= -\frac{1}{4}R\mathrm{_{0ikm}}x^m\gamma^{{\mathrm{i}}}\gamma^0 - \frac{1}{8} R\mathrm{_{jikm}} x^m \gamma^{{\mathrm{i}}}\gamma^{{\mathrm{j}}}.\nonumber
    \end{align}\c
    
    We thus obtain the following exact form of the Hamiltonian
    \begin{align}
        H = (g^{\tau\tau})^{-1}\gamma^\tau(i \gamma^i\partial_i+i \gamma^i \Gamma_i -m)\psi-i \Gamma_\tau,
    \end{align}
    or,to first order in curvature and acceleration, one can write
    \begin{align}
        H =& (g^{\tau\tau})^{-1}\gamma^\tau(i \gamma^{{\mathrm{i}}}\partial_i -m)\psi\\&- i \gamma^0\gamma^{I}(e^i_{I}-\delta^i_{I})\partial_i -i \gamma^0\gamma^{{\mathrm{i}}} \Gamma_i-i \Gamma_\tau.\nonumber
    \end{align}
    The only terms we are missing to expand the Hamiltonian are the expansions of $\gamma^\tau$ and $(g^{\tau\tau})^{-1}$. These can be easily done from the expansion of the frame and metric. We then obtain
    \begin{align}
        \gamma^\tau &= e^\tau_0 \gamma^0+ e^\tau_{{\mathrm{i}}} \gamma^{{\mathrm{i}}}\\& = \left(1-a_\mathrm{i} x^i -\frac{1}{2} R\mathrm{_{0k0m}}x^k x^m\right)\gamma^0-\frac{1}{6} R\mathrm{\indices{_{0l i m}}} x^{l} x^{m}\gamma^{{\mathrm{i}}}\nonumber
    \end{align}
    and
    \begin{align}
       g^{\tau \tau} &=  -1 + 2 a_\mathrm{i} x^i + R\mathrm{_{i 0 j 0}} x^i x^j,\\
       (g^{\tau \tau})^{-1} &=  -1 - 2 a_\mathrm{i} x^i - R\mathrm{_{i 0 j 0}} x^i x^j.
    \end{align}
    
    \begin{align}
        e_0^i-\delta_0^i &= \frac{1}{2}R\mathrm{\indices{^i_{l0m}}}x^lx^m\\
        e_{{\mathrm{i}}}^j-\delta_{{\mathrm{i}}}^j &= \frac{1}{6}R\mathrm{\indices{^j_{lim}}}x^lx^m
    \end{align}
    By plugging the individual expansions for the inverse metric component, $\gamma$ matrices and the Christoffel symbols, we obtain the expression   
    \begin{align}
        H =& (-1  -  a_\mathrm{i} x^i - \frac{1}{2}R\mathrm{_{k0m0}}x^kx^m )\gamma^0(i \gamma^{{\mathrm{i}}}\partial_i -m) \\
        &+\frac{1}{6} R\mathrm{\indices{_{0l j m}}} x^{l} x^{m}\gamma^{{\mathrm{j}}}(i \gamma^{{\mathrm{i}}}\partial_i -m) \nonumber\\
        &-\frac{i}{2}R\mathrm{\indices{^i_{l0m}}}x^lx^m\partial_i-\frac{i}{6}R\mathrm{\indices{^j_{lim}}}x^lx^m\gamma^0\gamma^{{\mathrm{i}}}\partial_j\nonumber\\
        &-i\gamma^0\gamma^{{\mathrm{i}}}\Gamma_i - i \Gamma_\tau\nonumber\\
        =& -\gamma^0(i \gamma^{{\mathrm{i}}}\partial_i -m)  \label{Vitamin}\\&-(  a_\mathrm{i} x^i + \frac{1}{2}R\mathrm{_{k0m0}}x^kx^m )\gamma^0(i \gamma^{{\mathrm{i}}}\partial_i -m)\nonumber \\ &+\frac{1}{6} R\mathrm{\indices{_{0l j m}}} x^{l} x^{m}\gamma^{{\mathrm{j}}}(i \gamma^{{\mathrm{i}}}\partial_i -m)\nonumber\\
        & -\frac{i}{2}R\mathrm{\indices{^i_{l0m}}}x^lx^m\partial_i-\frac{i}{6}R\mathrm{\indices{^j_{lim}}}x^lx^m\gamma^0\gamma^{{\mathrm{i}}}\partial_j\nonumber\\
        &+\frac{i}{4}R\mathrm{_{0ikm}}x^m\gamma^0\gamma^{{\mathrm{k}}}\gamma^{{\mathrm{i}}}\gamma^0 +\frac{i}{8} R\mathrm{_{jikm}} x^m \gamma^0\gamma^{{\mathrm{k}}}\gamma^{{\mathrm{i}}}\gamma^{{\mathrm{j}}}\nonumber\\
        &+\frac{i}{2}(a_\mathrm{i}+R\mathrm{_{0i0m}}x^m) \gamma^{{\mathrm{i}}}\gamma^0 \fix{-}\frac{i}{4}R\mathrm{_{0 m ij}}x^m\gamma^{{\mathrm{i}}}\gamma^{{\mathrm{j}}}.\nonumber
    \end{align}
    In the first line of the second equality we find the ``flat spacetime'' Hamiltonian. All the remaining terms are corrections due to acceleration of the curve or curvature. Notice that the bottom two rows are not in the simple shape presented in Section \ref{secDirWorldline}. To simplify the Hamiltonian we must use the following relations
    \begin{align}
        &R\mathrm{_{0ikm}}x^m\gamma^0\gamma^{{\mathrm{k}}}\gamma^{{\mathrm{i}}}\gamma^0 =\! -\!R\mathrm{_{0ikm}}x^m\gamma^0\gamma^{{\mathrm{k}}}\gamma^0\gamma^{{\mathrm{i}}}= R\mathrm{_{0ikm}}x^m\gamma^{{\mathrm{k}}}\gamma^{{\mathrm{i}}}\nonumber,\\
        &\b\gamma^{K} \gamma^{I} \gamma^{J}\!=\!\fix{-}\eta^{KI} \gamma^{J}\!\fix{-}\eta^{IJ} \gamma^{K}\!\fix{+}\eta^{K J} \gamma^{I}\!-\!i \epsilon^{L K I J} \gamma_{L} \gamma^{5},\c
    \end{align}
    \b where $\epsilon_{IJKL}$ is the Levi-Civita symbol in four dimensions. \c From which we obtain
    \begin{align}
        &R\mathrm{_{jikm}}x^m(\fix{-}\eta^{k i} \gamma^{j}\fix{-}\eta^{ij} \gamma^{k}-\eta^{k j} \gamma^{i}\fix{+}i \epsilon^{I k i j} \gamma_{I} \gamma^{5})\\
        &=R\mathrm{_{mkji}}x^m(\eta^{k i} \gamma^{j}-\eta^{k j} \gamma^{i}\fix{+}i \epsilon^{I k i j} \gamma_{I} \gamma^{5})\\
        &= R\mathrm{_{mkji}}x^m(\eta^{k i} \gamma^{j}-\eta^{k j} \gamma^{i})\\
        &=2R\mathrm{_{mkji}}x^m\eta^{k [i} \gamma^{j]}\\
        &=-2R\mathrm{_{mkij}}x^m\eta^{k i} \gamma^{j}
    \end{align}

    With this, the bottom two lines of Eq. \eqref{Vitamin} can then be rewritten as
    \begin{align}
        &\frac{i}{4}R\mathrm{_{0ikm}}x^m\gamma^0\gamma^{{\mathrm{k}}}\gamma^{{\mathrm{i}}}\gamma^0 +\frac{i}{8} R\mathrm{_{jikm}} x^m \gamma^0\gamma^{{\mathrm{k}}}\gamma^{{\mathrm{i}}}\gamma^{{\mathrm{j}}}\\
        &+\frac{i}{2}(a_\mathrm{i}+R\mathrm{_{0i0m}}x^m) \gamma^{{\mathrm{i}}}\gamma^0 \fix{-}\frac{i}{4}R\mathrm{_{0 m ij}}x^m\gamma^{{\mathrm{i}}}\gamma^{{\mathrm{j}}}\nonumber\\
        =& \frac{i}{4}R\mathrm{_{0ijm}}x^m\gamma^{{\mathrm{j}}}\gamma^{{\mathrm{i}}}+\frac{i}{4}R\mathrm{_{jikm}}x^m\delta^{k j}\gamma^0\gamma^{{\mathrm{i}}}\\
        & -\frac{i}{2} a_\mathrm{i} \gamma^0 \gamma^{{\mathrm{i}}}-\frac{i}{2} R\mathrm{_{0i0m}}x^m \gamma^0 \gamma^{{\mathrm{i}}}\fix{-}\frac{i}{4}R\mathrm{_{ ij0 m}}x^m\gamma^{{\mathrm{i}}}\gamma^{{\mathrm{j}}}\nonumber\\
        =& \left(-\frac{i}{2} a_\mathrm{i} -\frac{i}{2} R\mathrm{_{0i0m}}x^m+\frac{i}{4}R_{\mathrm{jikm}}x^m\delta^{k j}\right) \gamma^0 \gamma^{{\mathrm{i}}}\\
        & +\frac{i}{4}R\mathrm{_{0ijm}}x^m\gamma^{{\mathrm{j}}}\gamma^{{\mathrm{i}}}-\frac{i}{4}R\mathrm{_{ ij0 m}}x^m\gamma^{{\mathrm{i}}}\gamma^{{\mathrm{j}}}\nonumber\\
        =& \left(-\frac{i}{2} a_\mathrm{i} -\frac{i}{4} R\mathrm{_{0i0m}}x^m+\frac{i}{4}R\mathrm{_{im}}x^m\right) \gamma^0 \gamma^{{\mathrm{i}}}\\
        &  +\frac{i}{4}\left(R\mathrm{_{0jim}}x^m-R_{\mathrm{ij0m}}x^m\right)\gamma^{{\mathrm{i}}}\gamma^{{\mathrm{j}}}\nonumber\\
        =& -\frac{i}{4}\left(2a_\mathrm{i} + R\mathrm{_{0i0m}}x^m-R\mathrm{_{im}}x^m\right) \gamma^0 \gamma^{{\mathrm{i}}}\\
        & +\frac{i}{4}\left(R\mathrm{_{0jim}}x^m+R\mathrm{_{0mji}}x^m\right)\gamma^{{\mathrm{i}}}\gamma^{{\mathrm{j}}}\nonumber\\
        =& -\frac{i}{4}\left(2a_\mathrm{i} + R\mathrm{_{0i0m}}x^m-R\mathrm{_{im}}x^m\right) \gamma^0 \gamma^{{\mathrm{i}}}\\
        &-\frac{i}{4}R\mathrm{_{0imj}}x^m\gamma^{{\mathrm{i}}}\gamma^{{\mathrm{j}}}.\nonumber
    \end{align}
    This recovers the the expansion of Eq. \eqref{HDiracExp}.
    \color{black}
    %\begin{align}
        %-1 - 2 a_i x^i - R_{\alpha 0 \beta 0} x^\alpha x^\beta
    %\end{align}
    
    % \begin{align}
        %H &= (-1 - 2 a_i x^i - R_{\alpha 0 \beta 0} x^\alpha x^\beta)\left(1+a_i x^i +\frac{1}{2} R_{0k0m}x^k x^m\right)\gamma^0(i \gamma^{{\mathrm{i}}}\partial_i -m)-i\frac{1}{6} R\indices{^{\alpha}_{l i m}} x^{l} x^{m}\partial_i +i \gamma^{{\mathrm{i}}} \Gamma_i-i \Gamma_\tau
    %\end{align}
%\subsection{Checking that the acceleration terms are self-adjoint}
%\begin{align*}
        %(\phi,G\psi) &= \int \dd^3 x \:\phi^\dagger \left(- a_j x^j \gamma^0 \gamma^i i \partial_i - \frac{i}{2}a_i \gamma^0 \gamma^i \right)\psi\\
        %&= \int \dd^3 x \:\phi^\dagger \left(- a_j x^j (\gamma^0 \gamma^i)^\dagger i \partial_i\right)\psi +\phi^\dagger \left( \left(\frac{i}{2}a_i\right)^*(\gamma^0 \gamma^i)^\dagger \right)\psi\\
        %&= \int \dd^3 x \:(\gamma^0 \gamma^i\phi)^\dagger \left( ia_i  \right)\psi+(\gamma^0 \gamma^i\partial_i\phi)^\dagger \left( ia_jx^j  \right)\psi +\left(\frac{i}{2}a_i\gamma^0 \gamma^i\phi\right)^\dagger\psi\\
        %&= \int \dd^3 x \:(\gamma^0 \gamma^i\partial_i\phi)^\dagger \left( ia_i  \right)\psi+(-ia_jx^j\gamma^0 \gamma^i\phi)^\dagger \psi +\left(\frac{i}{2}a_i\gamma^0 \gamma^i\phi\right)^\dagger \psi\\
        %&= \int \dd^3 x \:(\gamma^0 \gamma^i\partial_i\phi)^\dagger \left( ia_i  \right)\psi+(-\frac{i}{2}a_jx^j\gamma^0 \gamma^i\phi)^\dagger \psi
    %\end{align*}
\onecolumngrid
\section{Reduction of Dirac Equation to Schr\"odinger Equation in Curved Spacetimes}\label{appSchCur}

For the reduction of Dirac's equation to the Schr\"odinger equation we proceed similarly to the approach taken in Section \ref{secSchAsDir} in flat spacetimes. First, we need to split the spinor into two two-component vectors and solve the bottom equation of motion for $\psi_B$. This can be done by getting the projection $P_B$ of the Hamiltonian and setting this equal to $i\partial_t\psi_B$. We will do so considering the minimal coupling with a $U(1)$ gauge field representing electromagnetism, so that we shift the partial derivative according to $\partial_i \mapsto D_i = \partial_i - i q A_i$ \fix{and $\partial_\tau \longmapsto D_\tau = \partial_\tau - i q A_0$}, where $q$ is the charge of the quantum system. With this in mind, the $B$ component of the equation of motion originated by the Hamiltonian from Eq. \eqref{HDiracExp} is found to yield
    \begin{align}
        &\left(i D_\tau + m + m a_\mathrm{j}x^j + \frac{m}{2}R\mathrm{_{k0m0}}x^kx^m  +  \frac{1}{6} R\mathrm{\indices{_{0l j m}}} x^{l} x^{m}\sigma^{{\mathrm{j}}}\sigma^{{\mathrm{i}}}i D_i  - \frac{i}{4}R\mathrm{_{0imj}}x^m\sigma^{{\mathrm{i}}}\sigma^{{\mathrm{j}}}+\frac{i}{2}R\mathrm{\indices{^i_{l0m}}}x^lx^m D_i\right)\psi_B\\
        &=\!\left( \!-i \sigma^{{\mathrm{i}}} D_i \!-\!  a_\mathrm{j} x^j \sigma^{{\mathrm{i}}}i D_i \!- \!\frac{1}{2}R\mathrm{_{k0m0}}x^kx^mi \sigma^{{\mathrm{i}}} D_i\! +\!\frac{m}{6} R\mathrm{\indices{_{0l j m}}} x^{l}x^m\sigma^{{\mathrm{j}}} \!-\! \frac{i}{6} R\mathrm{\indices{_{jl i m}}} x^{l} x^{m}\sigma^\mathrm{i} D^j \!-\! \frac{i}{4}\!\left(2a_\mathrm{i}\! +\! R\mathrm{_{0i0m}}x^m\!-\!R\mathrm{_{im}}x^m\right) \sigma^{{\mathrm{i}}}\!\right)\!\psi_A.\nonumber
    \end{align}
    This equation can be solved formally for $\psi_B$ in terms of $\psi_A$, similar to what was done in flat spacetimes in Eq. \eqref{approxFlat}. We are thus able to express the bottom component of the spinor $\psi$ in terms of the inverse of a differential operator acting on $H\psi_A$. This differential operator will now depend both on curvature and acceleration, as can be seen below    
    \begin{align}
        &\psi_B = \left(i  D_\tau + m + m a_\mathrm{j}x^j + \frac{m}{2}\mathrm{_{k0m0}}x^kx^m  +  \frac{1}{6} R\mathrm{\indices{_{0l j m}}} x^{l} x^{m}\sigma^{{\mathrm{j}}}\sigma^{{\mathrm{i}}}i D_i  - \frac{i}{4}R\mathrm{_{0imj}}x^m\sigma^{{\mathrm{i}}}\sigma^{{\mathrm{j}}}+\frac{i}{2}R\mathrm{\indices{^i_{l0m}}}x^lx^m D_i\right)^{-1}\nonumber\\
        &\left( \!-i \sigma^{{\mathrm{i}}} D_i\! -\!  a_\mathrm{j} x^j \sigma^{{\mathrm{i}}}i D_i \!-\! \frac{1}{2}R\mathrm{_{k0m0}}x^kx^mi \sigma^{{\mathrm{i}}} D_i \!+\!\frac{m}{6} R\mathrm{\indices{_{0l j m}}} x^{l}x^m\sigma^{{\mathrm{j}}}\! - \!\frac{i}{6} R\mathrm{\indices{_{jl i m}}} x^{l} x^{m}\sigma^\mathrm{i} D^j\! -\! \frac{i}{4}\left(2a_\mathrm{i} \!+\! R\mathrm{_{0i0m}}x^m\!-\!R\mathrm{_{im}}x^m\right) \sigma^{{\mathrm{i}}}\!\right)\!\psi_A.
    \end{align}
    
    The operator $i\partial_\tau$ corresponds to the total energy of the system. It is given by the rest mass of the system plus the non-relativistic energy, which can be associated with a differential operator\fix{s} $i \partial_T = i\partial_\tau - m$ \fix{ and $ D_T =\partial_T-i q A_0$}. We also factor out the term $1/2m$.     
    \begin{align}
        &\psi_B = \!\frac{1}{2m}\!\left( 1\!+\!\frac{i \partial_T\!+\! q A_0}{2m} \!+\! \frac{1}{2} a_\mathrm{j}x^j \!+\! \frac{1}{4}R\mathrm{_{k0m0}}x^kx^m  \!+\!  \frac{1}{12m} R\mathrm{\indices{_{0l j m}}} x^{l} x^{m}\sigma^{{\mathrm{j}}}\sigma^{{\mathrm{i}}}i D_i  \!-\! \frac{i}{8m}R\mathrm{_{0imj}}x^m\sigma^{{\mathrm{i}}}\sigma^{{\mathrm{j}}}\!+\!\frac{i}{4m}R\mathrm{\indices{^i_{l0m}}}x^lx^m D_i\right)^{-1}\nonumber\\
        &\left( \!-i \sigma^{{\mathrm{i}}} D_i \!- \! \b a_\mathrm{j} x^j \c  \sigma^{{\mathrm{i}}}i D_i \!-\! \frac{1}{2}R\mathrm{_{k0m0}}x^kx^mi \sigma^{{\mathrm{i}}} D_i \!+\!\frac{m}{6} R\mathrm{\indices{_{0l j m}}} x^{l}x^m\sigma^{{\mathrm{j}}} \!- \!\frac{i}{6} R\mathrm{\indices{_{jl i m}}} x^{l} x^{m}\sigma^\mathrm{i} D^j \!-\! \frac{i}{4}\left(2a_\mathrm{i}\! +\! R\mathrm{_{0i0m}}x^m\!-\!R\mathrm{_{im}}x^m\right) \sigma^{{\mathrm{i}}}\!\right)\!\psi_A.\label{C13}
    \end{align}
    Notice that the rest mass of the system tends to be much larger than any of the terms that show up in the expansion. In particular, it is usually larger than the non-relativistic energy, Coulomb potential, curvature and acceleration. This means that in a power expansion of the inverse operator in Eq. \eqref{C13}, it makes sense to neglect terms that will contribute with order $m^{-2}$. Therefore, the expansion of the inverse of the operator above up to first order in $m^{-1}$, curvature and acceleration yields
    \begin{align}
        \psi_B =& \left[\frac{1}{2m}\left(1 - \frac{1}{2} a_\mathrm{j}x^j - \frac{1}{4}R\mathrm{_{k0m0}}x^kx^m\right) + O\qty(m^{-2})\right]\\
        &\!\!\!\!\!\!\!\!\!\left( \!-i \sigma^{{\mathrm{i}}} D_i \!-\!  \b a_\mathrm{j} x^j \c  \sigma^{{\mathrm{i}}}i D_i \!-\! \frac{1}{2}R\mathrm{_{k0m0}}x^kx^mi \sigma^{{\mathrm{i}}} D_i \!+\!\frac{m}{6} R\mathrm{\indices{_{0l j m}}} x^{l}x^m\sigma^{{\mathrm{j}}} \!-\! \frac{i}{6} R\mathrm{\indices{_{jl i m}}} x^{l} x^{m}\sigma^\mathrm{i} D^j \!- \!\frac{i}{4}\left(2a_\mathrm{i} \!+\! R\mathrm{_{0i0m}}x^m\!-\!R\mathrm{_{im}}x^m\right) \sigma^{{\mathrm{i}}}\!\right)\!\psi_A\nonumber\\
        &\fix{+\frac{1}{24m}R_{0\mathrm{ljm}}x^lx^m\sigma^\mathrm{j}\,iD_T\psi_A}\\
        &\!\!\!\!\!\!\!\!\!\!\!\!= \frac{-1}{2m} \left(\! \left(1\! +\! \frac{1}{2} \b a_\mathrm{j} x^j \c   \!+\! \frac{1}{4}R\mathrm{_{k0m0}}x^kx^m\right)i \sigma^{{\mathrm{i}}} D_i \!-\!\frac{m}{6} R\mathrm{\indices{_{0l j m}}} x^{l}x^m\sigma^{{\mathrm{j}}}\!+\! \frac{i}{6} R\mathrm{\indices{_{jl i m}}} x^{l} x^{m}\sigma^\mathrm{i} D^j+ \frac{i}{4}\left(2a_\mathrm{i} \fix{+} R\mathrm{_{0i0m}}x^m \!-\! R\mathrm{_{im}}x^m\right)\! \sigma^{{\mathrm{i}}}\!\right)\!\psi_A\nonumber\\&\fix{+\frac{1}{24m}R_{0\mathrm{ljm}}x^lx^m\sigma^\mathrm{j}\,iD_T\psi_A}.\nonumber
    \end{align}%
    Notice that the few terms that come with a factor of $m$ in the calculations above are also first order in curvature. Thus, their product with any curvature or acceleration term can be neglected. We have thus been able to express $\psi_B$ in terms of a differential operator acting on $\psi_A$. For convenience\fix{, we define}
    \begin{align}
        \fix{D_B}&\fix{= \frac{-1}{2m} \left(\! \left(1\! +\! \frac{1}{2} a_\mathrm{j} x^j \!+\! \frac{1}{4}R\mathrm{_{k0m0}}x^kx^m\right)i \sigma^{{\mathrm{i}}} D_i \!-\!\frac{m}{6} R\mathrm{\indices{_{0l j m}}} x^{l}x^m\sigma^{{\mathrm{j}}}\!+\! \frac{i}{6} R\mathrm{\indices{_{jl i m}}} x^{l} x^{m}\sigma^\mathrm{i} D^j\! \fix{-} \frac{i}{4}\left(2a_\mathrm{i} \fix{+} R\mathrm{_{0i0m}}x^m \!-\! R\mathrm{_{im}}x^m\right)\! \sigma^{{\mathrm{i}}}\!\right),\nonumber}\\
        \fix{R_B} &\fix{=\frac{1}{24m}R_{0\mathrm{ljm}}x^lx^m\sigma^\mathrm{j},} 
    \end{align}
    \fix{so that we can write $\psi_B = (D_B + R_B\,iD_T )\psi_A$ to first order in curvature and acceleration.}% we call the operator from the Equation above $D_B$ so that we can write $\psi_B = D_B \psi_A$ to first order in curvature and acceleration.
    
    To proceed with the reduction of Dirac's formalism to Schr\"odinger's for the top component $\psi_A$ we write the equation of motion for $\psi_A$ as
    \begin{align}\label{C6}
        i\fix{D_\tau} \psi_A         &= P_A H \begin{pmatrix}\psi_A\\\psi_B\end{pmatrix} \equiv H_{AA}\psi_A + H_{AB}\psi_B = (H_{AA}+H_{AB}(D_B\fix{+R_{B}\,i D_T}))\psi_A \equiv \mathcal{H}_A\psi_A\fix{+R_{A} \, iD_T\psi_A},
    \end{align}
    where \fix{$R_A \equiv H_{AB}R_B$ and} $P_I$ is the projector in the $I$ component of the spinor for $I=A,B$ and $\mathcal{H}_A$ is the effective Hamiltonian for $\psi_A$. The Hamiltonian $H$ is then simply given by the expansion of Eq. \eqref{HDiracExp} and $H_{AA} \equiv P_A H P_A$, $H_{AB} \equiv P_A H P_B$. \fix{Moreover, it is possible to simplify Eq. \eqref{C6} to first order in curvature and acceleration. In fact, the term $R_A \,iD_T $ term can be written as}
    \begin{align}
        \fix{H_{AB}R_B \,iD_T\psi_A} &\fix{= -i \sigma^{{\mathrm{i}}} D_i\left(\frac{1}{24m}R_{0\mathrm{ljm}}x^lx^m\sigma^\mathrm{j}\,iD_T\psi_A\right)= \frac{-i}{24m}(R_{0\mathrm{ljm}}x^lx^mD_i+R_{0\mathrm{ijm}}x^m+R_{0\mathrm{lji}}x^l)\sigma^\mathrm{i}\sigma^\mathrm{j}(iD_\tau-m)\psi_A}
    \end{align}
    \fix{where we neglected derivatives of curvature and used $D_T = D_\tau - m$. Notice that the term above is of first order in curvature and in $1/m$. Then, we write}
    \begin{equation}
        \fix{\mathcal{H}_A = m - \frac{1}{2m}\sigma^{\textrm{i}}\sigma^\textrm{j}D_iD_j +h_A},
    \end{equation}
    \fix{where $h_A$ are the first order corrections due to curvature and acceleration. Then, Eq. \eqref{C6} can be written as}
    \begin{equation}
        \fix{(1-R_A)iD_\tau \psi_A = (1-R_A)m \psi_A- \frac{1}{2m}\sigma^{\textrm{i}}\sigma^\textrm{j}D_iD_j\psi_A + h_A\psi_A.}
    \end{equation}
    \fix{Notice that $R_A\frac{1}{2m}\sigma^{\textrm{i}}\sigma^\textrm{j}D_iD_j = \mathcal{O}(1/m^2)$ and $R_a h_A$ in of second order in our expansion in curvature and acceleration. That is, within the expansion we are considering, we can write}
    \begin{equation}
        \fix{(1-R_A)iD_\tau \psi_A = (1-R_A)\left(m \psi_A- \frac{1}{2m}\sigma^{\textrm{i}}\sigma^\textrm{j}D_iD_j\psi_A + h_A\psi_A\right) = (1-R_A)\mathcal{H}_A \psi_A,}
    \end{equation}
    \fix{so that to first order in curvature and acceleration, we obtain}
    \begin{equation}
        \fix{i D_\tau\psi_A = \mathcal{H}_A \psi_A \Leftrightarrow i \partial_\tau \psi_A = (\mathcal{H}_A - q A_0)\psi_A,}
    \end{equation}
    \fix{so that the effective Hamiltonian for the component $\psi_A$ is given by $\mathcal{H}_A - q A_0$.}
    
    This formulation by itself characterizes an approximation to Dirac's equation that recovers Schr\"odinger's equation for a two-component wavefunction $\psi_A$. However, we are interested in a Hamiltonian formulation for a complex wavefunction rather than for the two-component object $\psi_A$. We proceed by tracing out the spin degrees of freedom in the effective Hamiltonian $\mathcal{H}_A$. This means that Schr\"odinger's equation for a wavefunction $\psi$ then reads
    \begin{equation}
        i \partial_\tau \psi = \tr(\mathcal{H}_A)\psi.
    \end{equation}
    To obtain the Hamiltonian for the wavefunction it is then enough to calculate the trace of $H_{AA}$ and $H_{AB}D_B$ from the expansion of Eq. \eqref{C6}. The $H_{AA}$ term reads
    \begin{align}\label{HAA}
        \tr(H_{AA}) &= m +m a_\mathrm{j}x^j+\frac{m}{2}R\mathrm{_{k0m0}}x^kx^m  -\frac{1}{6} R\mathrm{\indices{_{0l i m}}} x^{l} x^{m}i D^i -\frac{i}{2}R\mathrm{\indices{^i_{l0m}}}x^lx^m D_i+\frac{i}{4}R\mathrm{_{m0}}x^m.
    \end{align}
    The $H_{AB}D_B$ term requires more work, since we must compute the composition of the operators $H_{AB} = P_A H P_B$ with $D_B$ to compute its trace. $H_{AB}$ then reads
    \begin{align}
        H_{AB} &= \! -i \sigma^{{\mathrm{i}}} D_i \! - \! a_\mathrm{j} x^j\sigma^{{\mathrm{i}}}i D_i \!-\! \frac{1}{2}R\mathrm{_{k0m0}}x^kx^m \sigma^{{\mathrm{i}}}i D_i  \!-\!\frac{m}{6} R\mathrm{\indices{_{0l j m}}} x^{l}x^m\sigma^{{\mathrm{j}}}\!-\!\frac{i}{6}R\mathrm{\indices{^j_{lim}}}x^lx^m\sigma^{{\mathrm{i}}} D_j\!-\! \frac{i}{4}\left(2a_\mathrm{i} \!+\! R\mathrm{_{0i0m}}x^m\!-\!R\mathrm{_{im}}x^m\right) \sigma^{{\mathrm{i}}}.
    \end{align}
    Its composition with $D_B$ is more easily computed when applied to the wavefunction $\psi_A$ due to the fact that it contains differential operators. %{\small
    %\begin{align*}
        %&H_{AB}D_B\psi_A \\
        %&= \frac{1}{2m}\Bigg(\Big(- \sigma^{{\mathrm{i}}}\sigma^{{\mathrm{j}}} D_i  D_j-i \sigma^{\bar{l}} D_l\Big(- \frac{1}{2} a_j x^j \sigma^{{\mathrm{i}}}i D_i - \frac{1}{4}R_{k0m0}x^kx^mi \sigma^{{\mathrm{i}}} D_i +\frac{m}{6} R\indices{_{0l j m}} x^{l}x^m\sigma^{{\mathrm{j}}} - \frac{i}{6} R\indices{_{jl i m}} x^{l} x^{m}\sigma^i D^j - \frac{i}{4}\left(2a_i + R_{0i0m}x^m-R_{im}x^m\right) \sigma^{{\mathrm{i}}}\Big)\\
        %&+\Big(-  a_i x^i\sigma^{{\mathrm{i}}}i D_i - \frac{1}{2}R_{k0m0}x^kx^m \sigma^{{\mathrm{i}}}i D_i  -\frac{m}{6} R\indices{_{0l j m}} x^{l}x^m\sigma^{{\mathrm{j}}}-\frac{i}{6}R\indices{^j_{lim}}x^lx^m\sigma^{{\mathrm{i}}} D_j- \frac{i}{4}\left(2a_i + R_{0i0m}x^m-R_{im}x^m\right) \sigma^{{\mathrm{i}}}\Big)(-i \sigma^{{\mathrm{j}}} D_j) \Bigg)\psi_A
    %\end{align*}}%
    Notice that as $D_i$ is a covariant derivative associated with the $U(1)$ gauge symmetry acting on the wavefunctions, it reduces to a partial derivative when applied to any function of $x$. The reason for the covariant derivative to act differently in complex wavefunctions is that $\psi_A$ is not a function but a section of a complex vector bundle associated with a $U(1)$ principal bundle. Taking this fact into consideration we obtain    \b
    \begin{align}
        H_{AB}D_B\psi_A =& \frac{1}{2m}\Bigg(- \sigma^{{\mathrm{i}}}\sigma^{{\mathrm{j}}} D_i  D_j -i \sigma^{\textrm{l}}D_l\Big(-\frac{i}{6}R\mathrm{\indices{^j_{kim}}}x^kx^m\sigma^{{\mathrm{i}}} D_j- \frac{i}{4}\left(2a_\mathrm{i} + R\mathrm{_{0i0m}}x^m-R\mathrm{_{im}}x^m\right) \sigma^{{\mathrm{i}}}\Big)\nonumber\\
        &+2\Big(-\frac{i}{6}R\mathrm{\indices{^j_{kim}}}x^kx^m\sigma^{{\mathrm{i}}} D_j- \frac{i}{4}\left(2a_\mathrm{i} + R\mathrm{_{0i0m}}x^m-R\mathrm{_{im}}x^m\right) \sigma^{{\mathrm{i}}}\Big)(-i\sigma_{\mathrm{l}}  D^l)-i\sigma^\mathrm{l} D_l\Big(\frac{m}{6} R\mathrm{\indices{_{0k j m}}} x^{k}x^m\sigma^{{\mathrm{j}}}\Big)\nonumber\\
        & -i \sigma^{\textrm{l}}D_l\Big(- \frac{1}{2} a_\mathrm{j} x^j \sigma^{{\mathrm{i}}}i D_{i} - \frac{1}{4}R\mathrm{_{k0m0}}x^kx^mi \sigma^{{\mathrm{i}}} D_i\Big)+\Big(-  a_\mathrm{j} x^j\sigma^{{\mathrm{i}}}i D_i - \frac{1}{2}R\mathrm{_{k0m0}}x^kx^m \sigma^{{\mathrm{i}}}i D_i \Big)(-i \sigma^{{\mathrm{l}}} D_l) \Bigg)\psi_A\nonumber\\
    %\begin{align*}
        %H_{AB}\psi_B =& \frac{1}{2m}\Big(- \sigma^{{\mathrm{i}}}\sigma^{{\mathrm{j}}} D_i  D_j -i \Big(-\frac{i}{6}R\indices{_{jlik}}x^l\sigma^{{\mathrm{k}}}\sigma^{{\mathrm{i}}} D^j-\frac{i}{6}R\indices{_{jkim}}x^m\sigma^{{\mathrm{k}}}\sigma^{{\mathrm{i}}} D^j- \frac{i}{4}\left( R_{0i0k}-R_{ik}\right)\sigma^{{\mathrm{k}}} \sigma^{{\mathrm{i}}}\Big)\\
        %&-\Big(\frac{1}{3}R\indices{^j_{lim}}x^lx^m\sigma^{{\mathrm{i}}}\sigma^{\textrm{l}}  D_l D_j+ \frac{1}{2}\left(2a_i + R_{0i0m}x^m-R_{im}x^m\right) \sigma^{{\mathrm{i}}}\sigma^{\textrm{l}}  D_l\Big) \\
        %&-i\frac{m}{6} R\indices{_{0l j k}} x^{l}\sigma^{\textrm{k}} \sigma^{{\mathrm{j}}}-i\frac{m}{6} R\indices{_{0k j m}} x^m\sigma^{\textrm{k}} \sigma^{{\mathrm{j}}}\\
        %& -i \sigma^{\bar{l}} D_l\Big( \frac{1}{2} a_j x^j  + \frac{1}{4}R_{k0m0}x^kx^m\Big)(-i\sigma^i  D_i)\\
        %&+\Big(a_i x^i + \frac{1}{2}R_{k0m0}x^kx^m  \Big)(-\sigma^{{\mathrm{i}}} \sigma^{{\mathrm{j}}} D_i D_j) \Big)\psi_A
    %\end{align*}
    %\textcolor{black}{The notation is a bit confusing because sometimes the derivative is acting only on the function after it and sometimes also on the wavefunction.}
    =& \frac{1}{2m}\Bigg(- \sigma^{{\mathrm{i}}}\sigma^{{\mathrm{j}}} D_i  D_j -\frac{1}{6}R\mathrm{\indices{_{jlik}}}x^l\sigma^{{\mathrm{k}}}\sigma^{{\mathrm{i}}} D^j-\frac{1}{6}R\mathrm{\indices{_{jkim}}}x^m\sigma^{{\mathrm{k}}}\sigma^{{\mathrm{i}}} D^j- \frac{1}{4}\left( R\mathrm{_{0i0k}}-R\mathrm{_{ik}}\right)\sigma^{{\mathrm{k}}} \sigma^{{\mathrm{i}}}\nonumber\\
    &-\Big(\frac{1}{3}R\mathrm{\indices{^j_{lim}}}x^lx^m\sigma^{{\mathrm{i}}}\sigma^{\textrm{k}}  D_k D_j+ \frac{1}{2}\left(2a_\mathrm{i} + R\mathrm{_{0i0m}}x^m-R\mathrm{_{im}}x^m\right) \sigma^{{\mathrm{i}}}\sigma^{\textrm{l}}  D_l\Big) \\
    &-i\frac{m}{6} R\mathrm{\indices{_{0l j k}}} x^{l}\sigma^{\textrm{k}} \sigma^{{\mathrm{j}}}-i\frac{m}{6} R\mathrm{\indices{_{0k j m}}} x^m\sigma^{\textrm{k}} \sigma^{{\mathrm{j}}} -\frac{1}{2}\Big(  a_\mathrm{j}   + R\mathrm{_{k0j0}}x^k\Big)\sigma^{\textrm{j}}\sigma^{\textrm{i}}  D_i\nonumber\\
    & -\Big( \frac{1}{2} a_\mathrm{k} x^k  + \frac{1}{4}R\mathrm{_{k0m0}}x^kx^m\Big)\sigma^{\textrm{j}}\sigma^{\textrm{i}}  D_j  D_i-\Big(a_\mathrm{k} x^k + \frac{1}{2}R\mathrm{_{k0m0}}x^kx^m  \Big)\sigma^{{\mathrm{i}}} \sigma^{{\mathrm{j}}} D_i D_j \Bigg)\psi_A.\nonumber
    \end{align} \c
    
    The expression above then allows one to identify the matrix valued differential operator $H_{AB}D_B$. Although this expression looks cumbersome, it is important to keep in mind that we are looking for the trace of such operator. To compute it we make use of the well known trace properties of sigma matrices (Eq. \eqref{traceProp}). We obtain
    %\begin{align*}
        %\tr (H_{AB}D_B) =& \frac{1}{2m}\Big(-  D_i  D^i -\frac{1}{6}R\indices{_{jkim}}x^m\delta^{ik} D^j+\frac{R}{4}\\
        %&-\Big(\frac{1}{3}R\indices{_{jlim}}x^lx^m  D^i D^j+ \frac{1}{2}\left(2a_i + R_{0i0m}x^m-R_{im}x^m\right)  D^i\Big) \\
        %&+i\frac{m}{6} R\indices{_{0 m}} x^m\\
        %& -\frac{1}{2}\Big(  a_i   + R_{k0i0}x^k\Big)  D^i\\
        %& -\Big( \frac{3}{2} a_j x^j  + \frac{3}{4}R_{k0m0}x^kx^m\Big) D^i  D_i \Big)
    %\end{align*}
    %\begin{align*}
        %\tr (H_{AB}D_B) =& \frac{1}{2m}\Big(-  D_i  D^i -\frac{1}{6}R\indices{_{jkim}}x^m\delta^{ik} D^j+\frac{R}{4}-\frac{1}{3}R\indices{_{jlim}}x^lx^m  D^i D^j-a_i D^i -\frac{1}{2}( R_{0i0m}-R_{im})x^m  D^i\\&+i\frac{m}{6} R\indices{_{0 m}} x^m -\frac{1}{2}\Big(  a_i   + R_{k0i0}x^k\Big)  D^i -\Big( \frac{3}{2} a_j x^j  + \frac{3}{4}R_{k0m0}x^kx^m\Big) D^i  D_i \Big)
    %\end{align*}
    \begin{align}
        \tr (H_{AB}D_B) =& -\frac{1}{2m}\left(1+\frac{3}{2} a_\mathrm{j} x^j  + \frac{3}{4}R\mathrm{_{k0m0}}x^kx^m\right)  D_i  D^i -\frac{1}{6m}R\mathrm{\indices{_{jlim}}}x^lx^m  D^i D^j\label{HABDB} \\
        &-\frac{1}{2m}\left(\frac{3}{2}a_\mathrm{i} +R\mathrm{_{0i0k}}x^k -\frac{1}{2}R\mathrm{_{ik}}x^k+\frac{1}{6}\delta^{jm}R\mathrm{_{imjk}}x^k\right) D^i+\frac{i}{12}R\mathrm{_{0k}}x^k +\frac{R}{8m}.\nonumber
    \end{align}

    We can then finally add together Eqs. \eqref{HAA} and \eqref{HABDB} in order to obtain the Hamiltonian $\mathcal{H}=\text{tr} (\mathcal{H}_A\fix{-q A_0})$ for the wavefunction $\psi$. We obtain
    \begin{align}
        \mathcal{H} =& \:m +m a_\mathrm{j}x^j+\frac{m}{2}R\mathrm{_{k0m0}}x^kx^m  -\frac{1}{6} R\mathrm{\indices{_{0l i m}}} x^{l} x^{m}i D^i -\frac{i}{2}R\mathrm{\indices{^i_{l0m}}}x^lx^m D_i+\frac{i}{4}R\mathrm{_{m0}}x^m\nonumber\\
        &-\frac{1}{2m}\left(1+\frac{3}{2} a_\mathrm{j} x^j  + \frac{3}{4}R\mathrm{_{k0m0}}x^kx^m\right)  D_i  D^i -\frac{1}{6m}R\mathrm{\indices{_{jlim}}}x^lx^m  D^i D^j \\
        &-\frac{1}{2m}\left(\frac{3}{2}a_\mathrm{i} +R\mathrm{_{0i0k}}x^k -\frac{1}{2}R\mathrm{_{ik}}x^k+\frac{1}{6}\delta^{jm}R\mathrm{_{imjk}}x^k\right) D^i+\frac{i}{12}R\mathrm{_{0k}}x^k +\frac{R}{8m}\fix{-qA_0}\nonumber%\\
        %=& \:m +m a_jx^j+\frac{m}{2}R_{k0m0}x^kx^m -\frac{1}{6} R\indices{_{0l i m}} x^{l} x^{m}i D^i -\frac{i}{2}R\indices{^i_{l0m}}x^lx^m D_i\\
        %&-\frac{1}{2m}\left(1+\frac{3}{2} a_j x^j  + \frac{3}{4}R_{k0m0}x^kx^m\right)  D_i  D^i -\frac{1}{6m}R\indices{_{jlim}}x^lx^m  D^i D^j \\
        %&-\frac{1}{2m}\left(\frac{3}{2}a_i +R_{0i0k}x^k -\frac{1}{2}R_{ik}x^k+\frac{1}{6}\delta^{jm}R_{imjk}x^k\right) D^i+\frac{i}{3}R_{0k}x^k +\frac{R}{8m}.
    \end{align}
    
    %The $a_i$-terms seem to be Hermitian. 
    
    %You can check that we can integrate $D_i$ as usual
    
    %\begin{equation}
        %\int \phi^* f(x) D_i \psi = \int \partial_i (\phi^* f(x) \psi) - \int D_i\triangleright\qty[\phi^* f(x)] \psi  = B.T. - \int (D_i \phi)^*f(x) \psi - \int \phi^* \partial_i f(x) \psi
    %\end{equation}

\section{Hermiticity of the Hamiltonian}\label{appHermitian}
    
    In this appendix we show that the Hamiltonian $\mathcal{H}$ from Eq. \eqref{YES} is Hermitian with respect to the inner product defined for the space of complex wavefunctions associated with the spacelike hypersurfaces $\Sigma_\tau$ (Eq. \eqref{innerProdSigmaExp}) up to first order in curvature and acceleration. In order to do this we will explicitly integrate each term in $(\phi,H\psi)$ by parts and show that we obtain the same differential operator $H$ acting on the function $\phi$. We will do so when the theory is coupled to electromagnetism, so we use the fact that the derivatives $D_\mu = \partial_\mu - iq A_\mu$ allow for integration by parts just like the $\partial_\mu$ operators.
    
    We start by showing that the terms that are a function of the acceleration are a Hermitian operator. Namely, we will start by showing that the following operator is Hermitian in the expansion we are considering,
    \begin{align}
        A =& m a_\mathrm{j}x^j-\frac{3}{4m} a_\mathrm{j} x^j D_iD^i-\frac{3}{4m}a_\mathrm{i}D^i.
    \end{align}
    Notice the operator $A$ is already of first order in the expansion we are considering. This means that for the purpose of showing Hermitianness of this term we only need to consider the flat inner product from Eq. \eqref{innerProdSigmaFlat}. With this in mind, we compute $(\phi,A\psi)$ for two arbitrary wavefunctions $\phi$ and $\psi$. We obtain
    \begin{align}
        &\int \dd[3]{x} \phi^*\left( m a_\mathrm{j}x^j-\frac{3}{4m} a_\mathrm{j} x^j  D_i D^i - \frac{3}{4m}a_\mathrm{i} D^i \right)\psi\nonumber\\
        &= \int \dd[3]{x} \left(m a_\mathrm{j} x^j + \frac{3}{4m} a_\mathrm{i}  D^i\right) \phi^*\psi - \frac{3}{4m}  D_i  D^i\left( a_\mathrm{j} x^j \phi^*\right) \psi \nonumber\\
        &= \int \dd[3]{x} \left(m a_\mathrm{j} x^j + \frac{3}{4m} a_\mathrm{i}  D^i\right) \phi^*\psi - \frac{3}{4m}  D_i \left( a^\mathrm{i} \phi^* + a_\mathrm{j} x^j  D^i \phi^*\right) \psi \nonumber\\
        &= \int \dd[3]{x} \left(m a_\mathrm{j} x^j + \frac{3}{4m} a_\mathrm{i}  D^i\right) \phi^*\psi - \frac{3}{2m} a^\mathrm{i}  D_i \phi^* \psi - \frac{3}{4m} a_\mathrm{j} x^j  D_i  D^i \phi^*\psi \nonumber\\
        &= \int \dd[3]{x} \left(m a_\mathrm{j} x^j - \frac{3}{4m} a_\mathrm{i}  D^i  - \frac{3}{4m} a_\mathrm{j} x^j  D_i  D^i\right) \phi^*\psi = (A\phi,\psi),
    \end{align}
    which is enough to show that the acceleration terms are Hermitian. 
    
    The next step is then to show that the Hamiltonian is Hermitian without the presence of acceleration. That is, we consider the operator
    \begin{align}
        \mathcal{H} =& \textcolor{black}{-\frac{1}{2m}\left(1  + \frac{3}{4}R\mathrm{_{k0m0}}x^kx^m\right)  D_i  D^i} \textcolor{black}{-\frac{1}{6m}R\mathrm{\indices{_{jlim}}}x^lx^m  D^i D^j}\label{Ha=0} \\
        &\textcolor{black}{-\frac{1}{2m}\left(R\mathrm{_{0i0k}}x^k -\frac{1}{2}R\mathrm{_{ik}}x^k+\frac{1}{6}\delta^{jm}R_{\mathrm{{imjk}}}x^k\right) D^i}+\frac{i}{3}R\mathrm{_{0k}}x^k\textcolor{black}{-\frac{2}{3}R\mathrm{\indices{_{0mil}}}x^lx^m i D^i},\nonumber
    \end{align}
    which corresponds precisely to the Hamiltonian of Eq. \eqref{YES} when one sets $a_i = 0$. We will use the technique of integration by parts in each one of the terms above separately in other to obtain their adjoints and then add these together in order to calculate the adjoint of the operator $\mathcal{H}$. Notice that the only term that contains a factor which is not proportional to curvature is the kinetic term. Therefore, it will also be the only term that will require us to take into account the correction to the inner product due to curvature. Integrating by parts we obtain
    \begin{align}
        &\int \dd^3 x \phi^*(x)\left(1-\frac{1}{6}\left(R_{0\mathrm{i}0\mathrm{j}} + R_{\mathrm{i}\mathrm{j}}\right)x^ix^j\right)\left(1  + \frac{3}{4}R\mathrm{_{k0m0}}x^kx^m\right)  D_i  D^i\psi(x)\nonumber\\
        =&\int \dd^3 x  D_k  D^k\Big(\phi^*(x)\left(1-\frac{1}{6}\left(R_{0\mathrm{i}0\mathrm{j}} + R_{\mathrm{i}\mathrm{j}}\right)x^ix^j\right)\left(1  + \frac{3}{4}R\mathrm{_{k0m0}}x^kx^m\right)\Big) \psi(x)\nonumber\\
        =&\int \dd^3 x  D_k \Big( D^k\phi^*(x)\left(1-\frac{1}{6}\left(R_{0\mathrm{i}0\mathrm{j}} + R_{\mathrm{i}\mathrm{j}}\right)x^ix^j+ \frac{3}{4}R\mathrm{_{k0m0}}x^kx^m\right)\Big)  \psi(x)\nonumber\\
        &+\int \dd^3 x  D_k \Big( \phi^*(x)\partial^k\left(1-\frac{1}{6}\left(R_{0\mathrm{i}0\mathrm{j}} + R_{\mathrm{i}\mathrm{j}}\right)x^ix^j+ \frac{3}{4}R\mathrm{_{k0m0}}x^kx^m\right)\Big)  \psi(x)\nonumber\\
        =&\int \dd^3 x  D_k  D^k\phi^*(x)\left(1-\frac{1}{6}\left(R_{0\mathrm{i}0\mathrm{j}} + R_{\mathrm{i}\mathrm{j}}\right)x^ix^j+ \frac{3}{4}R\mathrm{_{k0m0}}x^kx^m\right) \psi(x)\nonumber\\
        &+2\int \dd^3 x  D_k\phi^*(x)\partial^k\left(1-\frac{1}{6}\left(R_{0\mathrm{i}0\mathrm{j}} + R_{\mathrm{i}\mathrm{j}}\right)x^ix^j+ \frac{3}{4}\b R\mathrm{_{k0m0}}\c x^kx^m\right) \psi(x)\nonumber\\
        &+\int \dd^3 x \phi^*(x)\partial_k\partial^k\left(1-\frac{1}{6}\left(R_{0\mathrm{i}0\mathrm{j}} + R_{\mathrm{i}\mathrm{j}}\right)x^ix^j  + \frac{3}{4}R\mathrm{_{k0m0}}x^kx^m\right)  \psi(x)        \label{blue}\\
        =&\int \dd^3 x  D_i  D^i\phi^*(x)\left(1-\frac{1}{6}\left(R_{0\mathrm{i}0\mathrm{j}} + R_{ij}\right)x^ix^j\right)\left(1  + \frac{3}{4}R\mathrm{_{k0m0}}x^kx^m\right)  \psi(x)\nonumber\\
        &+2\int \dd^3 x  D_k\phi^*(x)\left(-\frac{1}{3}\left(R_{0\mathrm{k}0\mathrm{j}} + R_{\mathrm{k}\mathrm{j}}\right)x^j+ \frac{3}{2}R\mathrm{_{k0m0}}x^m\right)  \psi(x)\nonumber\\
        &+\int \dd^3 x \phi^*(x)\partial_k\partial^k\left(1-\frac{1}{3}\left(R_{0\mathrm{i}0\mathrm{j}} + R_{\mathrm{i}\mathrm{j}}\right)\delta^{ij}  + \frac{3}{2}R\mathrm{_{i0j0}}\delta^{ij}\right)  \psi(x)\nonumber\\
        =&\int \dd^3 x  D_i  D^i\phi^*(x)\left(1-\frac{1}{6}\left(R_{0\mathrm{i}0\mathrm{j}} + R_{ij}\right)x^ix^j\right)\left(1  + \frac{3}{4}R\mathrm{_{k0m0}}x^kx^m\right)  \psi(x)\nonumber\\
        &\textcolor{black}{+2\int \dd^3 x D^i\phi^*(x)\left(-\frac{1}{3}R\mathrm{_{ij}} + \frac{7}{6}R\mathrm{_{0i0j}}\right)x^j \psi(x)}\nonumber\\
        &\textcolor{black}{+\int \dd^3 x \phi^*(x)\left(-\frac{1}{3}\delta^{ij}R\mathrm{_{ij}} + \frac{7}{6}R\mathrm{_{00}}\right) \psi(x)},\nonumber
    \end{align}
    where the first term in the last equality above is exactly the operator itself acting in the wavefunction $\phi$ instead of $\psi$. The adjoint of this operator then includes the two extra terms from the bottom two lines of Eq. \eqref{blue}. Because none of the terms that constitute the Hamiltonian of Eq. \eqref{Ha=0} is Hermitian,  we will see a similar pattern for the remaining computations of this section.
    
    The second term from Eq. \eqref{Ha=0}, when applied to two wavefunctions can be recast as
    \begin{align}
       &\int \dd^3 x \phi^*(x)R\mathrm{\indices{_{jlim}}}x^lx^m  D^i D^j \psi(x) 
       = \int \dd^3 x  D^i D^j\Big(\phi^*(x)R\mathrm{\indices{_{jlim}}}x^lx^m\Big)\psi(x)\nonumber\\
       &= \int \dd^3 x  D^i\left( D^j\phi^*(x)R\mathrm{\indices{_{jlim}}}x^lx^m\right)\psi(x)+\int \dd^3 x  D^i\left( \phi^*(x)D^j\Big(R\mathrm{\indices{_{jlim}}}x^lx^m\Big)\right)\psi(x)\nonumber
       \\
       &= \int \dd^3 x  D^i\Big( D^j\phi^*(x)R\mathrm{\indices{_{jlim}}}x^lx^m\Big)\psi(x)+\int \dd^3 x  D^i\Big( \phi^*(x)\Big(R\mathrm{\indices{_{jlim}}}(\delta^{lj}x^m + \delta^{mj}x^l)\Big)\Big)\psi(x)\nonumber\\
       &= \int \dd^3 x  D^i\Big( D^j\phi^*(x)R\mathrm{\indices{_{jlim}}}x^lx^m\Big)\psi(x)+\int \dd^3 x  D^i\Big( \phi^*(x)\delta^{mj}R\mathrm{\indices{_{jlim}}}x^l\Big)\psi(x)\nonumber\\
       &= \int \dd^3 x  D^i D^j\phi^*(x)R\mathrm{\indices{_{jlim}}}x^lx^m\psi(x)+\int \dd^3 x  D^j\phi^*(x) R\mathrm{\indices{_{jlim}}}\delta^{li}x^m\psi(x)\nonumber\\
       &+\int \dd^3 x  D^i \phi^*(x)\delta^{mj}R\mathrm{\indices{_{jlim}}}x^l\psi(x)+\int \dd^3 x   \phi^*(x)\delta^{mj}R\mathrm{\indices{_{jlim}}}\delta^{li}\psi(x)\nonumber\\
       &= \int \dd^3 x  D^i D^j\phi^*(x)R\mathrm{\indices{_{jlim}}}x^lx^m\psi(x)\textcolor{black}{+2\int \dd^3 x  D^i \phi^*(x)\delta^{mj}R\mathrm{\indices{_{jlim}}}x^l\psi(x)}\textcolor{black}{+\int \dd^3 x   \phi^*(x)\delta^{mj}R\mathrm{\indices{_{jlim}}}\delta^{li}\psi(x)}.\label{red}
    \end{align}
    We can again identify two additional terms after computing the adjoint of such operator. The third term from Eq. \eqref{Ha=0} yields  
    \begin{align}
        &\int \dd^3 x \phi^*(x) \left(R\mathrm{_{0i0k}}x^k -\frac{1}{2}R\mathrm{_{ik}}x^k+\frac{1}{6}\delta^{jm}R\mathrm{_{imjk}}x^k\right) D^i \psi(x)\label{orange}\\
        &= \!-\!\int \dd^3 x D^i \phi^*(x) \left(\!R\mathrm{_{0i0k}}x^k \!-\!\frac{1}{2}R\mathrm{_{ik}}x^k\!+\!\frac{1}{6}\delta^{jm}R\mathrm{_{imjk}}x^k\!\right)\! \psi(x)\!-\! \int \!\dd^3 x \phi^*(x) D^i \left(\!R\mathrm{_{0i0k}}x^k \!-\!\frac{1}{2}R\mathrm{_{ik}}x^k\!+\!\frac{1}{6}\delta^{jm}R\mathrm{_{imjk}}x^k\!\right)\! \psi(x)\nonumber\\
        &= \textcolor{black}{\!-\!\int \!\dd^3 x D^i \phi^*(x) \left(\!R\mathrm{_{0i0k}}x^k \!-\!\frac{1}{2}R\mathrm{_{ik}}x^k\!+\!\frac{1}{6}\delta^{jm}R\mathrm{_{imjk}}x^k\right)\! \psi(x)}\textcolor{black}{-\! \int \!\dd^3 x \phi^*(x)  \left(\!R_{00} \!-\!\frac{1}{2}R\mathrm{_{ik}}\delta^{ki}\!+\!\frac{1}{6}\delta^{ki}\delta^{jm}R\mathrm{_{imjk}}\!\right)\! \psi(x)}.\nonumber
    \end{align}
    The fourth term in  Eq. \eqref{Ha=0} is simply a product term, so that its adjoint is simply complex conjugation. Given that the fourth term is multiplication by a purely imaginary function, it is anti-Hermitian. Finally, the adjoint of the last term in Eq. \eqref{Ha=0} yields
    \begin{align}
        &\int \dd^3 x \phi^*(x)R\mathrm{\indices{^i_{l0m}}}x^lx^m D_i \psi(x)\nonumber\\
        & -\int \dd^3 x D_i\phi^*(x)R\mathrm{\indices{^i_{l0m}}}x^lx^m  \psi(x) -\int \dd^3 x \phi^*(x)R\mathrm{\indices{^i_{l0m}}}D_ix^lx^m  \psi(x)\nonumber\\
        & -\int \dd^3 x D_i\phi^*(x)R\mathrm{\indices{^i_{l0m}}}x^lx^m  \psi(x) -\int \dd^3 x \phi^*(x)(R\mathrm{\indices{^i_{i0k}}}x^k +R\mathrm{\indices{^i_{k0i}}}x^k ) \psi(x)\nonumber\\
        & -\int \dd^3 x D_i\phi^*(x)R\mathrm{\indices{^i_{l0m}}}x^lx^m  \psi(x) \textcolor{black}{+\int \dd^3 x \phi^*(x)R\mathrm{\indices{_{0k}}}x^k \psi(x)}.\label{green}
    \end{align}
    
    We thus have all the terms that constitute the adjoint of the Hamiltonian when $a_i = 0$. We then define the operator $K$ by the following equation
    \begin{align}\label{OhhNoH}
        (\phi,\mathcal{H}\psi) \equiv (\mathcal{H}\phi,\psi) + (K \phi,\psi)_0.
    \end{align}
    Then we have that $\mathcal{H}$ is Hermitian if and only if $K=0$. Out of Eqs. \eqref{blue}, \eqref{red},\eqref{orange} and \eqref{green} we find $K$ to be given by
    \begin{align}
        K =& \textcolor{black}{-\frac{1}{2m}\left(-\frac{1}{3}\delta^{ij}R\mathrm{_{ij}} + \frac{7}{6}R_{00}\right)}-\textcolor{black}{\frac{1}{6m}\delta^{mj} \delta^{li}R\mathrm{\indices{_{jlim}}}}+\textcolor{black}{\frac{1}{2m}\left(R_{00} -\frac{1}{2}R\mathrm{_{ik}}\delta^{ki}+\frac{1}{6}\delta^{ki}\delta^{jm}R\mathrm{_{imjk}}\right)}\nonumber\\
        &+\textcolor{black}{\frac{1}{m}\left(R\mathrm{_{0i0k}}x^k -\frac{1}{2}R\mathrm{_{ik}}x^k+\frac{1}{6}\delta^{jm}R\mathrm{_{imjk}}x^k\right)D^i}+\frac{2i}{3}R\mathrm{_{0k}}x^k- \textcolor{black}{\frac{2i}{3}R\mathrm{\indices{_{0k}}}x^k} \\
        &\textcolor{black}{-\frac{1}{m}\left(-\frac{1}{3}R\mathrm{_{ij}} + \frac{7}{6}R\mathrm{_{0i0j}}\right)x^jD_i}-\textcolor{black}{\frac{1}{3m}\delta^{mj}R\mathrm{\indices{_{jlim}}}\b x^lD^i\c }.\nonumber
    \end{align}
    %\begin{align}
        %LOL=& \textcolor{black}{\frac{1}{6m}\delta^{ij}R_{ij}-\frac{7}{12m}R_{00}}-\frac{1}{6m}\delta^{mj} \delta^{li}R\indices{_{jlim}}+\textcolor{black}{\frac{1}{2m}R_{00}-\frac{1}{4m}R_{ik}\delta^{ki}+\frac{1}{12m}\delta^{ki}\delta^{jm}R_{imjk}}\\
        %&+\frac{1}{m}R_{0i0k}x^kD^i -\frac{1}{2m}R_{ik}x^kD^i+\frac{1}{6m}\delta^{jm}R_{imjk}x^kD^i+\frac{2i}{3}R_{0k}x^k- \frac{2i}{3}R\indices{_{0k}}x^k \\
        %&\frac{1}{3m}R_{ij}x^jD_i- \frac{7}{6m}R_{0i0j}x^jD_i-\frac{1}{3m}\delta^{mj}R\indices{_{jlim}}x^lD^i
    %\end{align}
    We can then regroup the similar terms together, obtaining
    \begin{align}
        K = -\frac{1}{6m}(R\mathrm{_{0i0k}}+R_{ik}+\delta^{jm}R\mathrm{_{imjk}})\left(\frac{1}{2}\delta^{ki}+x^kD^i\right).
    \end{align}
    %\begin{align}
        %LOL=& \left(\textcolor{black}{\frac{1}{2m}}-\frac{7}{12m}\right)R_{00}+\left(\frac{2i}{3}- \frac{2i}{3}\right)R\indices{_{0k}}x^k +\left(\frac{1}{6m}\textcolor{black}{-\frac{1}{4m}}\right)\delta^{ij}R_{ij} +\left(-\frac{1}{6m}\textcolor{black}{+\frac{1}{12m}}\right)\delta^{mj}\delta^{li}R\indices{_{jlim}}\\
        %&+\left(\frac{1}{m}-\frac{7}{6m}\right)R_{0i0k}x^kD^i + \left(\frac{1}{3m}-\frac{1}{2m}\right)R_{ik}x^kD^i+\left(\frac{1}{6m}-\frac{1}{3m}\right)\delta^{jm}R_{imjk}x^kD^i
    %\end{align}
    %\begin{align}
        %LOL=& -\frac{1}{12m}R_{00} -\frac{5}{12m}\delta^{ij}R_{ij} -\frac{1}{12m}\delta^{mj}\delta^{li}R\indices{_{jlim}}-\frac{1}{6m}R_{0i0k}x^kD^i -\frac{1}{6m}R_{ik}x^kD^i-\frac{1}{6m}\delta^{jm}R_{imjk}x^kD^i\\
        %=& -\frac{1}{12m}R_{00} -\frac{5}{12m}\delta^{ij}R_{ij} -\frac{1}{12m}\delta^{mj}\delta^{li}R\indices{_{jlim}}-\frac{1}{6m}\left(R_{0i0k} +R_{ik}+\delta^{jm}R_{imjk}\right)x^kD^i
    %\end{align}
    %\begin{align}
        %&-\frac{1}{12m}\delta^{ij}R_{0i0j} -\frac{1}{12m}\delta^{ij}(\delta^{km}R_{kimj}-R_{0i0j}) -\frac{1}{12m}\delta^{ij}\delta^{km}R\indices{_{jkmi}}\\
        %&=-\frac{1}{12m}\delta^{ij}\delta^{km}R_{kimj} +\frac{1}{12m}\delta^{ij}\delta^{km}R\indices{_{kimj}}=0
    %\end{align}
    It is then possible to show that the first factor is zero by decomposing the Ricci tensor in terms of the time and space components of the Riemann tensor,
    \begin{align}
        R\mathrm{_{0i0j}} +R\mathrm{_{ij}}+\delta^{km}R\mathrm{_{imkj}} &=  R\mathrm{_{0i0j}} +(\delta^{km}R\mathrm{_{kimj}}-R\mathrm{_{0i0j}})+\delta^{km}R\mathrm{_{imkj}}\nonumber\\
        &= \delta^{km}(R\mathrm{_{kimj}}+R\mathrm{_{imkj}}) =  \delta^{km}(R\mathrm{_{kimj}}-R\mathrm{_{mikj}}) =0.
    \end{align}
    Therefore we have shown that $K=0$, so that Eq. \eqref{OhhNoH} yields
    \begin{equation}
        (\phi,\mathcal{H}\psi) = (\mathcal{H}\phi,\psi).
    \end{equation}
    That is, the Hamiltonian found for the system is Hermitian with respect to the rest surface's inner product to first order.

\bibliography{references}
    
\end{document}